\documentclass[pra,twocolumn,superscriptaddress,nofootinbib]{revtex4-2}
\usepackage{amsfonts}
\usepackage{amsmath}
\usepackage{bbm}
\usepackage{amssymb}
\usepackage{graphicx}%
\usepackage{footmisc}
\usepackage{bm}
\usepackage{braket}
\usepackage{CJK}
\usepackage{hyperref}
\usepackage{cleveref}
\usepackage{longtable,booktabs}
\usepackage[all,arc,knot,matrix]{xy}

\crefname{equation}{Eq.}{Eqs.}

\newcommand{\abs}[1]{\lvert#1\rvert}
\newcommand{\norm}[1]{\lVert#1\rVert}
\newcommand{\Tr}{\text{Tr}}

\begin{document}
\begin{CJK*}{UTF8}{gbsn}
\title{Integral fluctuation theorems and trace-preserving map}

\author{Zhiqiang Huang (黄志强)}
\email{hzq@wipm.ac.cn}
\affiliation{ Innovation Academy for Precision Measurement Science and Technology, Chinese Academy of Sciences, Wuhan 430071, China}

\date{\today}

\begin{abstract}
    The detailed fluctuation theorem implies symmetry in the generating function of entropy production probability. The integral fluctuation theorem directly follows from this symmetry and the normalization of the probability. In this paper, we rewrite the generating function by integrating measurements and evolution into a constructed mapping. This mapping is completely positive, and the original integral FT is determined by the trace-preserving property of these constructed maps. We illustrate the convenience of this method by discussing the eigenstate fluctuation theorem and heat exchange between two baths. This set of methods is also applicable to the generating functions of quasi-probability, where we observe the Petz recovery map arising naturally from this approach.
\end{abstract}


\maketitle

\section{Introduction}\label{INTRO}
As the generalized second law of thermodynamics from a microscopic perspective, the fluctuation theorem (FT) is of fundamental importance in non-equilibrium statistical mechanics. The FT focuses mainly on the irreversibility of entropy production. While a general FT can be given under unitary evolution \cite{EHM09}, it is only useful when the entropy production can be expressed in terms of physical and measurable quantities, which usually requires the initial state of the bath to be the canonical distribution. However, when this condition is not met, the FT will deviate. By calculating the characteristic (generating) function of entropy production, \cite{IKS17,IKS22} proved that the integral FT approximation holds in both the long and short time regimes, even when the initial state of the bath is a single energy eigenstate of many-body systems. The proof of this result is complex and uses various forms of the eigenstate thermalization hypothesis (ETH), which prevents further development of the approaches used therein and makes it difficult to tighten the errors. The difficulty in estimating deviation stems from the lack of a unified framework to study the impact of initial state deviation, as the generating function also includes two-point measurements and evolution, each of which can have additional effects based on the initial state deviation. The forms of these operators are quite different in conventional quantum mechanics formalism, which makes deviation estimation even more difficult.

The integral FT is directly related to the normalization of the backward processes, which in turn depends on the trace preservation (TP) property of the backward map. In this sense, the generating function should be determined by the properties of some mapping. The modified propagator approach is one such method \cite{EHM09}. It incorporates measurements and Fourier transforms into the modified propagator. Here we propose a different approach. We start with the general FT under unitary evolution and incorporate measurements and Fourier transforms into quantum operations. These quantum operations, along with the evolution, form a complete positive (CP) map. If this constructed map is trace preserving when $\lambda=i$, then the generating function $G(i)=1$ and the integral FT holds. Since the generating function is just rewritten here, it is equivalent to the original FT. However, by unifying the measurements and evolution into the constructed mapping, considering the impact of initial state deviation becomes simpler and more unified. To illustrate this advantage, we first consider the general quantum lattice model in \cite{IKS17,IKS22}. Our aim is to provide a streamlined proof of the eigenstate FT, relying on fewer assumptions (specifically, the weak ETH assumption rather than several strong versions of ETH). After that, we consider another important class of FTs concerning the heat exchange between two baths \cite{JW04}. We will investigate whether two baths, whose initial states consist of single-energy eigenstates, also exhibit FT.

The exploration of FTs through the construction of CP mappings has long been proposed by \cite{MHP15, AMOP16}. These studies primarily delve into the properties of the evolution of quantum open systems. Their approaches are closely connected to the FTs for the quantum channel \cite{MHP18,KK19}, which has established another general framework of quantum FTs. Using the Petz recovery map, it provides the detailed FT for two-point measurement quasiprobabilities. Notably, the resulting FT does not incorporate measurements of the environment. This distinction highlights a fundamental difference from the entropy production considered in the FT under unitary evolution. However, since the quantum channel can emerge naturally from the open system formalism, there likely exists a relationship between the quantum channel FTs and the unitary evolution FTs.  If the global unitary operation $U$ has a global fixed point, \cite{LP21} demonstrates that both methods can yield the same entropy production. Correspondingly, by employing the approach outlined in this paper, we demonstrate that the Petz recovery map can be naturally derived by analyzing the generating function of quasi-probabilities. This offers an alternative perspective on the relationship between the two methods.

The paper is organized as follows: In \cref{S2}, we first briefly introduce the operator-state formalism. Then, we rewrite the generating function with complete positive mappings and briefly discuss the properties of the constructed mappings. We delve into the core component $\mathcal{N}_\beta(t)$ of the constructed mapping in detail in \cref{PPMB}. In \cref{S3} and \cref{S4}, we utilize the constructed mapping to discuss the eigenstate FT. The integral FT is shown to hold within a small margin of error, and a detailed calculation of error estimation can be found in \cref{EESEC}. In \cref{S5}, we discuss the integral FT for the quantum channel and uncover the significance of the corresponding constructed mapping. Section \ref{S6} concludes.

\section{The integral fluctuation theorem and trace-preserving map}\label{S2}
\subsection{Preliminaries}
Since we are mainly concerned with CP mappings in this paper, we use the superoperator form for simplicity. Furthermore, the FT necessitates the consideration of inverse mappings and measurements. These components have simpler expressions within the operator-state formalism, which we also employ. It's noteworthy that these formalisms maintain mathematical equivalence with the conventional framework of quantum mechanics.

Let us first introduce the operator state  $|O)$ for an operator $O$ on the Hilbert space $\mathcal{H}_S$ of quantum states of a system $S$, which belongs to a new  Hilbert space $\mathsf{H}$ with an inner product defined by 
\begin{equation}
    (O_1|O_2):=\text{Tr}(O_1^\dag O_2).
\end{equation}
  An orthonormal basis for $\mathsf{H}$ is $|\Pi_{ij})$ where $\Pi_{ij}=\ket{i}\bra{j}$, as one can check that 
  \begin{equation}
  (\Pi_{kl}|\Pi_{ij})=\delta_{ik}\delta_{jl}.
  \end{equation}
  The completeness of this basis $\{|\Pi_{ij})\}$ is
    \begin{equation}\label{3}
        \mathcal{I}=\sum_{ij}|\Pi_{ij})(\Pi_{ij}|.
  \end{equation}
  If only the diagonal elements are summed, the corresponding superoperator is actually a dephasing map
  \begin{equation}
    \sum_{i}|\Pi_{i})(\Pi_{i}|\rho)=| \sum_{i}\ket{i}\bra{i}\rho\ket{i}\bra{i}).
  \end{equation} 
The conjugate relation reads
\begin{align}\label{c4}
    (O_1|O_2)^\dagger:=\text{Tr}(O_1^\dag O_2)^\dagger=\text{Tr}(O_2^\dag O_1)\notag\\
    =(O_2|O_1)=(O_1^\dagger|O_2^\dagger).
\end{align}
The unitary superoperator acting on the operator state gives
\begin{equation}\label{DFUSPO}
    \mathcal{U}|O):=|U O U^\dagger).
\end{equation}
The conjugate relation shows that
\begin{align}\label{CJRSPO}
    (O_1|\mathcal{U}|O_2)^\dagger=\text{Tr}(O_1 U O_2^\dag U^\dagger ) =(O_1^\dag|\mathcal{U}|O_2^\dag)\notag \\
    =\text{Tr}(O^\dagger_2 U^\dagger O_1 U  )= (O_2 |\mathcal{U}^\dagger |O_1).
\end{align}
The trace of an operator $\rho$ can be given by $(I|\rho)$.
The superoperator $\mathcal{N}$ is trace-preserving if and only if it satisfies $ (I|\mathcal{N}=(I|$.

Now we briefly introduce the FT. The FT can be framed in a unifying language in terms of entropy production \cite{LP21}. The entropy production $\Delta\sigma$ is generally related to thermodynamic quantities such as Gibbs-von~Neumann entropy, work, heat, and free energy. Its formulation depends on the underlying physical system. In detailed FT (Crooks FT), the forward entropy production is exponentially more likely than the reverse 
\begin{equation}
    \frac{P_F(\Delta\sigma)}{P_B(-\Delta\sigma)}=e^{\Delta\sigma},
\end{equation}
where $P_F$ is the forward probability distribution of entropy production and $P_B$ is the backward one. They are given by the probability density function of the difference between two measurements
\begin{equation}
    P_F(\Delta\sigma)=\sum_{\sigma_t,\sigma_0}\delta(\Delta\sigma -(\sigma_t-\sigma_0))P_F(\sigma_t,\sigma_0).
\end{equation}
Its generating function 
\begin{equation}
    G_F(\lambda):=\int_{-\infty}^{\infty}d \Delta\sigma e^{i\lambda \Delta\sigma}P_F(\Delta\sigma)
\end{equation}
is often used. The parameter $\lambda$ is a complex number, and people are generally concerned about the properties of the generating function when $\lambda=i$. The Crooks FT implies the fundamental symmetry on the generating function
\begin{equation}
    G_F(\lambda)=\int_{-\infty}^{\infty}d \Delta\sigma e^{i\lambda \Delta\sigma+\Delta\sigma}P_B(-\Delta\sigma)=G_B(i-\lambda).
\end{equation}
An integral FT immediately follows from the normalization of $P_B$
\begin{equation}\label{IFT}
    \braket{e^{-\Delta\sigma}}= G_F(i)= G_B(0)=1.
\end{equation}
Combining it with Jensen's inequality $\braket{e^{-X}}\geq e^{-\braket{X}}$, we can get a generalization of the second law of thermodynamics 
\begin{equation}
    \braket{\Delta\sigma}\geq - \ln \braket{e^{-X}}=0.
\end{equation}

\subsection{Constructs a completely positive map from the generating function}
In this section, we explore how to construct a CP map from the generating function and connect the FT with the TP property of this constructed CP map.

Consider an isolated, possible driven, quantum system evolves according to the unitary evolution. 
The probability density function of the difference between two measurements can be expressed as
\begin{equation}\label{JPTPMFP}
    P_F(\Delta\sigma)=\sum_{\sigma_t,\sigma_0}\delta(\Delta\sigma -(\sigma_t-\sigma_0)) ( \Pi_{\sigma_t}|\mathcal{U}| \Pi_{\sigma_0})( \Pi_{\sigma_0}|\rho_0),
\end{equation}
where $\rho_0$ is the initial density matrices for the forward process. The observable $\hat{\sigma}_t$ is a Hermitian operator.
The eigenvalues (eigenbasis) of $\hat{\sigma}_t$ are denoted by $\sigma_t$ ($\Pi_{\sigma_t}=\ket{\sigma_t}\bra{\sigma_t}$): $\hat{\sigma}_t=\sum_{\sigma_t} \sigma_t\Pi_{\sigma_t} $. The generating function of probability (\ref{JPTPMFP}) is
\begin{equation}\label{GFPT1}
    G_F(\lambda):=\sum_{\sigma_0,\sigma_t} (e^{i\lambda(\hat{\sigma}_t-\hat{\sigma}_0)}|\Pi_{\sigma_t}\otimes \Pi_{\sigma_0})( \Pi_{\sigma_t}|\mathcal{U}| \Pi_{\sigma_0})( \Pi_{\sigma_0}|\rho_0).
\end{equation}
The factor $e^{-i\lambda\hat{\sigma}}$ rescaling the measurement results indeed. It is related to the rescaling map $\mathcal{J}^{\alpha}_{O}(\cdot):=O^\alpha (\cdot)O^{\alpha\dagger}$, where $O$ is arbitrary operator and $\alpha$ is the power index. \emph{Here we only consider the cases where the real part of $\lambda$ is zero}, and then we have
\begin{align}
    \mathcal{J}_{e^{- \hat{\sigma}_0}}^{i\lambda/2} | \Pi_{\sigma_0})=|e^{-i\lambda\hat{\sigma}_0/2}\Pi_{\sigma_0} e^{-i\lambda\hat{\sigma}_0/2})\notag\\
    = e^{-i\lambda\sigma_0} | \Pi_{\sigma_0})=(e^{-i\lambda\hat{\sigma}_0}| \Pi_{\sigma_0})| \Pi_{\sigma_0}).
\end{align}
Using it, we can rewrite some items in the generating function as follows
\begin{equation}\label{EXP2RSM}
    \sum_{\sigma_0}  (e^{-i\lambda\hat{\sigma}_0}| \Pi_{\sigma_0})| \Pi_{\sigma_0})( \Pi_{\sigma_0}| =\mathcal{J}_{e^{- \hat{\sigma}_0}}^{i\lambda/2} \circ\mathcal{M}_{\hat{\sigma}_0},
\end{equation}
 where $\mathcal{M}_{\hat{\sigma}_0}:=\sum_{\sigma_0} | \Pi_{\sigma_0})( \Pi_{\sigma_0}|$ is a dephasing map. The rescaling map is CP, but generally not TP. It is worth pointing out that the rescaling map here is very similar to the re-weighting approach used in \cite{HMSSS22}, where this approach is used to prepare the thermal state.  With \cref{EXP2RSM}, we can rewrite \cref{GFPT1} as
\begin{equation}\label{GFRSM}
    G_F(\lambda)=(I|\mathcal{J}_{e^{- \hat{\sigma}_t}}^{-i\lambda/2}\circ \mathcal{M}_{\hat{\sigma}_t}\circ\mathcal{U}(t)\circ\mathcal{J}_{e^{- \hat{\sigma}_0}}^{i\lambda/2}\circ \mathcal{M}_{\hat{\sigma}_0}|\rho_0)
\end{equation}
or its conjugated form
\begin{equation}\label{GFRSMCF}
    G_F(\lambda)=(I|\mathcal{J}_{\rho_0}^{1/2} \circ \mathcal{M}_{\hat{\sigma}_0}\circ\mathcal{J}_{e^{- \hat{\sigma}_0}}^{i\lambda/2}\circ \mathcal{U}^\dagger(t)|e^{i \lambda \hat{\sigma}_t})^\dagger
\end{equation}
Where $\mathcal{M}_{\hat{\sigma}_t}$ is omitted because it shares the same basis as $e^{i \lambda \hat{\sigma}_t}$. Additionally, we omitted the $\dag$ on $\mathcal{M}$ and $\mathcal{J}_{e^{- \hat{\sigma}}}$ because they are self-conjugate. If $\{\Pi_{\sigma_0}\}$ is the diagonal basis of $\rho_0$, then $\mathcal{M}_{\hat{\sigma}_0}$ can also be omitted. In \cref{GFRSM,GFRSMCF}, the generating function is completely rewritten with the operator state and completely positive maps. As will be shown later, the integral FT will be determined by the TP property of these maps.

Take $\hat{\sigma}_t=-\ln \hat{\rho}_t$, then we have
\begin{equation}\label{IDQS}
    e^{-i\lambda \hat{\sigma}_0}=(\hat{\rho}_0)^{i\lambda}.
\end{equation}
In such cases, the $\mathcal{M}_{\hat{\sigma}_0}$ can be omitted
\begin{equation}\label{RFMDC}
    \mathcal{J}_{e^{- \hat{\sigma}_0}}^{i\lambda/2} \circ\mathcal{M}_{\hat{\sigma}_0}|\rho_0)=\mathcal{J}_{e^{- \hat{\sigma}_0}}^{i\lambda/2}|\rho_0).
\end{equation}
Combining \cref{GFRSM,RFMDC}, we get
\begin{align}
    G_F(i)=(I|\mathcal{J}_{e^{- \hat{\sigma}_t}}^{1/2}\circ\mathcal{U}\circ\mathcal{J}_{e^{- \hat{\sigma}_0}}^{-1/2} | \rho_0)
    =(\rho_t|\mathcal{U}(t)|I)=1.
\end{align}
Therefore, for a closed system, the entropy production $\ln (\hat{\rho}_0/\hat{\rho}_t)$ satisfies integral FT. Since $\braket{-\ln \hat{\rho}_t}=S(\rho_t)$, the average of this entropy production is the same as the change of the Gibbs-von~Neumann entropy of the system. 

For all cases in this paper, we have $\Tr e^{- \hat{\sigma}_t}=1$, which can also be achieved in other cases  by adding a normalization factor. Under this condition, according to \cref{GFRSMCF}, the integral FT is satisfied if and only if the following constructed CP map is also a TP map
\begin{equation}\label{CTRSPO1}
    \mathcal{J}_{\rho_0}^{1/2} \circ \mathcal{M}_{\hat{\sigma}_0}\circ\mathcal{J}_{e^{- \hat{\sigma}_0}}^{-1/2}\circ \mathcal{U}^\dagger(t).
\end{equation}
This map provides a new perspective for considering the integral FT. The mapping (\ref{CTRSPO1}) is not necessarily a TP mapping, it depends on the initial state $\rho_0$ as well as the initial measurements $\hat{\sigma}_0$. If observable are taken as \cref{IDQS}, it is easy to prove that the constructed map (\ref{CTRSPO1}) is TP
\begin{equation}
    (I|\mathcal{J}_{\rho_0}^{1/2} \circ \mathcal{M}_{\hat{\sigma}_0}\circ\mathcal{J}_{e^{- \hat{\sigma}_0}}^{-1/2}\circ \mathcal{U}^\dagger(t)=(I|,
\end{equation}
which is consistent with the previous conclusion that the integral FT holds.

\subsection{The constructed map for the open quantum system}
Now consider an open quantum system $S$ interacting with an environment $E$; $S$ and $E$ form a closed system with unitary evolution. If there is no correlation between the initial system and the environment, we can set $\hat{\sigma}_t=-\ln \hat{\rho}_S(t)-\ln \hat{\rho}_E$, and then we have
\begin{equation}\label{NCSE}
    e^{-i\lambda \hat{\sigma}_0}=\hat{\rho}^{i\lambda}_S(0)\otimes\hat{\rho}^{i\lambda}_E.
\end{equation}
Similar to the case of closed systems, it is easy to prove that open systems also satisfy integral FT
\begin{equation}\label{TFT}
    G_F(i)=(I|\mathcal{U}^\dagger_{SE}(t)|\hat{\rho}_S(t)\otimes \hat{\rho}_E)^\dagger=1.
\end{equation}
This integral FT is completely general, but only contains quantities like entropy, which require knowledge of all information about the state of the environment. In open quantum systems, the environment is usually considered to be relatively large and obtaining all information about it is difficult. Therefore, it is necessary to introduce some thermodynamic quantities such as work, free energy, and heat to simplify the measurement. For example, it is usually assumed that the initial state of the heat bath is the canonical ensemble $\hat{\rho}_E=e^{-\beta \hat{H}_E}/Z_E$, so we can set
\begin{equation}\label{TSL}
    \hat{\sigma}_t=\beta (\hat{H}_E- F_E)-\ln \hat{\rho}_S(t).
\end{equation}
If we make no assumptions about the initial state of the environment, but still set $\hat{\sigma}_t$ as \cref{TSL}, then we have
\begin{align}\label{EIFT}
    G_F(\lambda)=(I_{SE}|\mathcal{J}_{\rho_S(t)\otimes\rho^\text{can}_E}^{-i\lambda/2}\circ \mathcal{U}_{SE}(t)\notag \\
    \circ\mathcal{J}_{\rho_S(0)\otimes \rho^\text{can}_E}^{i\lambda/2}\circ \mathcal{M}_{\hat{\sigma}_0}^E|\rho_{S}(0)\otimes\rho_{E}(0)).
\end{align}
Since the environment $\rho_E(0)$ can deviate from the canonical ensemble $\rho^\text{can}_E$, the integral FT will have errors and $ G_F(i)$ can deviate from 1. 

\begin{figure}[t]
    \centering
    \includegraphics[width=0.5\textwidth]{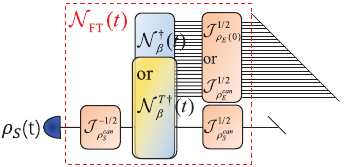}
    \caption{A quantum circuit diagram for the generating function \cref{IFTDT}. In this rewritten form, the terms related to the operator expansion are grouped as $ \mathcal{N}^\dag_\beta(t)$. According to the Lieb-Robinson bound, only part of the environment effectively participates in the expansion. Therefore, $\mathcal{N}^\dag_\beta(t)$ can be replaced by $\mathcal{N}^{T\dag}_\beta(t)$ with only a small error. After the replacement, according to the ETH, the total environment energy eigenstate is also indistinguishable from the canonical ensemble, so $\rho_{E}(0)$ can be replaced by $\rho_S^\text{can}$.  As shown in  \cref{EFNFT}, a TP mapping can be obtained after several replacements.  Deviations from the integral FT can be derived by considering the errors caused by these replacements.
    }\label{FIG1}
    \end{figure}

The deviation of the environment state is not a sufficient condition for the violation of the integral FT, which is also related to the system-environment interaction. For non-driven systems, the Hamiltonian of the whole system can generally be written as $H=H_S+H_E+H_I$. If there is no system-environment interaction $H_I=0$, then the overall generating function can be separated into a system part and an environment part $ G_F(\lambda)= G^S_F(\lambda) G^E_F(\lambda)$. Since the system state does not deviate, we have $G^S_F(i)=1$. And the environment part gives
\begin{equation}
    G^E_F(\lambda)=(I_E|\mathcal{J}_{\rho^\text{can}_E}^{-i\lambda/2}\circ \mathcal{U}_{E}(t)
    \circ\mathcal{J}_{ \rho^\text{can}_E}^{i\lambda/2}\circ \mathcal{M}_{\hat{\sigma}_0}^E|\rho_{E}(0)).
\end{equation}
Since $[U_E(t),\rho^\text{can}_E]=0$, there is always $ G^E_F(i)=1$, no matter whether $\rho_{E}(0)$ deviates from $\rho^\text{can}_E$ or not. This inspires us to decompose the rescaling map in \cref{EIFT} as follows
\begin{equation}\label{RSMDC}
    \mathcal{J}_{\rho_S(t)\otimes \rho^\text{can}_E}=\mathcal{J}_{\rho^\text{can}_S\otimes \rho^\text{can}_E}\circ\mathcal{J}^{-1}_{\rho^\text{can}_S}\circ\mathcal{J}_{\rho_S(t)}.
\end{equation}
$\mathcal{J}^{-1}_{\rho^\text{can}_S}\circ\mathcal{J}_{\rho_S(t)}$ is a local superoperator of the system. Using decomposition (\ref{RSMDC}), we can rewrite the generating function (\ref{EIFT}) as
\begin{align}\label{LMEV}
    G_F(\lambda)=(I_{SE}|\mathcal{J}_{\rho_S(t)}^{-i\lambda/2}\circ\mathcal{J}_{\rho_S^\text{can}}^{i\lambda/2}\circ\mathcal{N}_{\beta}(t)\notag \\
    \circ\mathcal{J}_{\rho_S^\text{can}}^{-i\lambda/2}\circ\mathcal{J}_{\rho_S(0)}^{i\lambda/2}|\rho_{S}(0)\otimes\rho'_{E}(0)),
\end{align}
where 
\begin{equation}\label{SOFNB}
    \mathcal{N}_\beta(t)(\cdot):= \mathcal{J}^{-i\lambda /2}_{\rho^\text{can}_S\otimes \rho^\text{can}_E}\circ\mathcal{U}_{SE}(t)\circ \mathcal{J}^{i\lambda /2}_{\rho^\text{can}_S\otimes \rho^\text{can}_E}(\cdot),
\end{equation}
and $\rho'_{E}(0)=\mathcal{M}_{\hat{\sigma}_0}^E[\rho_{E}(0)]$ is the environment density matrix resulting from the decoherence of the environment due to the measurement.

Similar to \cref{CTRSPO1},  the conjugated form of \cref{LMEV} gives \footnote{In \cref{IFTDT}, we omit the $I_{E}$ in the right parenthesis, which will change the meaning of the first $\mathcal{J}_{\rho_{E}}^{1/2}$ in $ \mathcal{N}_{\text{FT}}$: it becomes an assignment map $\rho_S\to \rho_S\otimes \rho_{E}$, which is linear and CPTP. The similar approaches are also used in \cref{S4}.}
\begin{equation}\label{IFTDT}
     G_F(i)=( I_{SE}| \mathcal{N}_{\text{FT}}(t)|\rho_S(t))^\dagger,
\end{equation}
and the integral FT holds if and only if the following constructed map is a TP map
\begin{equation}\label{CSTCPM}
   \mathcal{N}_{\text{FT}}(t):= \mathcal{J}_{\rho'_{E}(0)}^{1/2} \circ \mathcal{J}_{\rho_S^\text{can}}^{1/2}\circ \mathcal{N}_\beta^\dagger(t)\circ \mathcal{J}^{-1/2}_{\rho^\text{can}_S}.
\end{equation}
The key parts of the formulae (\ref{CSTCPM}) are the mapping $\mathcal{N}_\beta(t)$ related to the evolution and the mapping $ \mathcal{J}_{\rho'_{E}(0)}^{1/2}$ related to the initial state and the initial measurement. If the initial state of the environment is the canonical ensemble, then we have $\mathcal{N}_{\text{FT}}(t)=\mathcal{U}^\dagger_{SE}(t)\circ \mathcal{J}_{\rho^\text{can}_{E}}^{1/2}$. It acts on the operator state and gives  $\mathcal{N}_{\text{FT}}(t)|\rho_S(t))=\mathcal{U}^\dagger_{SE}(t)|\rho_S(t)\otimes \rho^\text{can}_{E})$. Therefore, it is indeed a TP map in this case. When the environment $\rho_E(0)$ deviates from the canonical ensemble $\rho^\text{can}_E$, the constructed map $\mathcal{N}_{\text{FT}}(t)$ is generally not a TP map. The integral FT will have errors and $ G_F(i)$ can deviate from 1. The inverse temperature in \cref{CSTCPM} remains arbitrary, allowing for optimization of the deviation estimate by selecting it based on the state of the environment. For instance, when considering an environment $X$ initialized in a single-energy eigenstate $\ket{E^X_a}$, the choice of inverse temperature is typically determined by aligning the total energy, often set as $\beta_a$ such that
\begin{equation}\label{ITFE}
   E_a=(e^{-\beta_a H_X}/Z_{\beta_a}|H_X). 
\end{equation}

\section{Eigenstate fluctuation theorem}\label{S3}
In this section, we will discuss the properties of $\mathcal{N}_{\text{FT}}(t)$ and the FT when the initial state of the environment is the energy eigenstate. Since $\rho_{E}(0)$ is the energy eigenstate of $\hat{H}_E$, we have $\rho'_{E}(0)=\rho_{E}(0)$. Our proof draws on the idea in article \cite{IKS22} about using Lieb-Robinson bound for short-time regime and time averaging for long-time regime. But the approach is very different: we mainly consider the deviation of the generating function through the properties of the mapping $\mathcal{N}_{\text{FT}}(t)$. It is worth pointing out that if we replace the $\mathcal{N}_{\beta}(t)$ in \cref{IFTDT} with $\mathcal{U}_{SE}(t)$, then we will get the $ \delta G_S(i)$ in \cite{IKS22}. Therefore, $\mathcal{N}_{\beta}(t)-\mathcal{U}_{SE}(t)$ is directly related to the interaction-induced error $\delta G_I$ in \cite{IKS22}.

The underlying physical insight behind the proof is as follows: The error of the integral FT is determined by the interaction between the system and the environment. According to the Lieb-Robinson bound \cite{HHKL21} or the operator growth hypothesis \cite{PCASA19}, the influence range of the interaction is limited under the (imaginary) time evolution. In the short-time regime, the expanded operator is still unable to distinguish the pure state of the many-body systems from the canonical ensemble, according to the ETH. In the long-time regime, the expanded energy width is limited, and the expanded operator is unable to distinguish these states according to the ETH.

\subsection{Short-time regime}\label{STR}
When the evolution time tends to zero, it is obvious that $ \mathcal{N}_{\text{FT}}(t)= \mathcal{J}_{\rho_{E}(0)}^{1/2}$. It acts on the operator state and gives $\mathcal{N}_{\text{FT}}(t)|\rho_S(t))=|\rho_S(t)\otimes \rho_{E}(0))$. Therefore, it is a TP map and the FT holds.
If the evolution time is short, it is foreseeable that the error can be bounded, as we will discuss in detail below.

When the evolution time is short, the two measurements about the system can only be influenced by a small piece of environment due to the limitation of information flow speed. Considering the ETH, it is difficult to distinguish the overall pure state of the environment from the canonical bath in this small area. This makes the integral FT approximately valid. From the perspective of constructed map $ \mathcal{N}_{\text{FT}}$, we can rewrite it as 
\begin{align}\label{EFNFT}
   \mathcal{N}_{\text{FT}}(t) 
    = \mathcal{J}_{\rho_{E}(0)}^{1/2} \circ \mathcal{J}_{\rho_S^\text{can}}^{1/2}\circ [\mathcal{N}_\beta^\dagger(t)-\mathcal{N}_\beta^{T\dagger}(t)]\circ \mathcal{J}^{-1/2}_{\rho^\text{can}_S}\notag\\
    +[\mathcal{J}_{\rho_{E}(0)}^{1/2}-\mathcal{J}_{\rho^\text{can}_E}^{1/2}] \circ \mathcal{J}_{\rho_S^\text{can}}^{1/2}\circ \mathcal{N}_\beta^{T\dagger}(t)\circ \mathcal{J}^{-1/2}_{\rho^\text{can}_S}\notag \\
    +\mathcal{J}_{\rho^\text{can}_E}^{1/2} \circ \mathcal{J}_{\rho_S^\text{can}}^{1/2}\circ [\mathcal{N}_\beta^{T\dagger}(t)-\mathcal{N}_\beta^\dagger(t)]\circ \mathcal{J}^{-1/2}_{\rho^\text{can}_S} \notag\\
    +\mathcal{J}_{\rho^\text{can}_E}^{1/2} \circ \mathcal{J}_{\rho_S^\text{can}}^{1/2}\circ \mathcal{N}_\beta^\dagger(t)\circ \mathcal{J}^{-1/2}_{\rho^\text{can}_S} ,
\end{align}
where $\mathcal{N}_\beta^{T}(t)$ is defined in \cref{STETS}. It is a truncated version of $\mathcal{N}_\beta(t)$ based on the speed limit of information flow. The last term of formula (\ref{EFNFT}) is indeed a TP mapping. Therefore, the deviation of the integral FT can be estimated using the bounds of other terms. The terms in the first and third lines of \cref{EFNFT} represent the error due to truncation, which can be bounded by the Lieb-Robinson bound. See \cref{ETSTR} for specific calculations. The term in the second line of \cref{EFNFT} represents the error caused by the deviation of the initial state. Since the mapping $\mathcal{J}_{\rho_S^\text{can}}^{1/2}\circ \mathcal{N}_\beta^{T\dagger}(t)\circ \mathcal{J}^{-1/2}_{\rho^\text{can}_S}$ acts on state $|\rho_S(t))$ and gives a local operator, this part of the error can be bounded by ETH. See \cref{EIDES} for details. Finally, in the large-environment limit, by combining the error terms in \cref{EFNFT}, we obtain
\begin{align}\label{EBEFNEF}
    \abs{\delta G_F(i)}\leq \abs{ G_F(i)-G^T_F(i)}+\abs{G^T_F(i)-G^{\text{can},T}_F(i)} \notag \\
    +\abs{G^{\text{can},T}_F(i)-G^\text{can}_F(i)} =o(1),
\end{align}
where the difference in the generating function corresponds one-to-one with the error terms in \cref{EFNFT}. The superscript "$\text{can}$" represents replacing the initial environment state in $\mathcal{N}_{\text{FT}}(t)$ with the canonical ensemble, while "$T$" represents replacing $\mathcal{N}_\beta$ in $\mathcal{N}_{\text{FT}}(t)$ with $\mathcal{N}^T_\beta$. We illustrate these replacements in \cref{FIG1}. From the figure, we can observe that when using the truncated mapping $\mathcal{N}_\beta^{T}(t)$, only a portion of the environment will participate, which is why ETH can be applied.

\subsection{Long-time regime}\label{LTA}
When the evolution time is long, the system can gather sufficient information about the environment, allowing it to distinguish the pure state environment from the thermal state environment. Consequently, under the long-term evolution, the integral FT may exhibit significant  deviations. However, while large deviations are possible,  they are typically transient.  This transient nature stems from the system eventually reaching equilibrium as the evolution time lengthens. For most times, the total system tends to approach a fixed steady state \cite{GLMSW17}. Considering the ETH, distinguishing between these steady states with local observations becomes challenging. As a result, the integral FT remains approximately valid in the long-term average.

 For any quantity $O(t)$, its long-time average is defined as follows
\begin{equation}
    \overline{O(t)}:=\lim_{T\to \infty}\frac{1}{T}\int_0^{T} O(t)dt.
\end{equation}
In the generating function, not only does $\mathcal{N}_{\text{FT}}(t)$ change with time, but the final state of the system also evolves over time
    \begin{equation}
        |\rho_S(t))=\mathcal{E}_t |\rho_S(0))=(I_E|\mathcal{U}_{SE}(t)|\rho_S(0)\otimes\rho_E(0)).
    \end{equation}
    Combining it with \cref{IFTDT,EFNFT}, we have
    \begin{align}\label{CSMLAVG}
        \overline{\mathcal{N}_{\text{FT}}(t)\mathcal{E}_t}
         = \overline{[\mathcal{N}_{\text{FT}}(t)-\mathcal{N}^{T'}_{\text{FT}}(t)]\circ \mathcal{E}_t}\notag\\
         + [\mathcal{J}_{\rho_{E}(0)}^{1/2}-\mathcal{J}_{\rho^\text{can}_E}^{1/2}] \circ \mathcal{J}_{\rho_S^\text{can}}^{1/2}\circ  \overline{\mathcal{N}_\beta^{T'\dagger}(t)\circ \mathcal{J}^{-1/2}_{\rho^\text{can}_S}\circ \mathcal{E}_t}\notag \\
         +\mathcal{J}_{\rho^\text{can}_E}^{1/2} \circ \mathcal{J}_{\rho_S^\text{can}}^{1/2}\circ \overline{[\mathcal{N}_\beta^{T'\dagger}(t)-\mathcal{N}_\beta^\dagger(t)]\circ \mathcal{J}^{-1/2}_{\rho^\text{can}_S}\circ \mathcal{E}_t }\notag\\
         +\mathcal{J}_{\rho^\text{can}_E}^{1/2} \circ \mathcal{J}_{\rho_S^\text{can}}^{1/2}\circ \overline{\mathcal{N}_\beta^\dagger(t)\circ \mathcal{J}^{-1/2}_{\rho^\text{can}_S}\circ \mathcal{E}_t} ,
     \end{align}
     where $\mathcal{N}_\beta^{T'}(t)$ is defined in \cref{TMLTR}, representing another truncated version of $\mathcal{N}_\beta(t)$ based on the limit imposed by information flow speed under the imaginary time evolution. By substituting $\mathcal{N}_\beta$ with $\mathcal{N}_\beta^{T'}$ in $\mathcal{N}_{\text{FT}}$, we obtain $\mathcal{N}^{T'}_{\text{FT}}$. The last term of \cref{CSMLAVG} is equivalent to $\overline{\mathcal{U}^\dagger_{SE}(t)\circ \mathcal{J}_{\rho^\text{can}_{E}}^{1/2}\circ \mathcal{E}_t}$, which is also a TP mapping. The terms in the first and third lines of \cref{CSMLAVG} represent the error due to truncation, which can be constrained by the Lieb-Robinson bound. Detailed calculations are provided in \cref{ETSTR}. Under long-time averaging, a single unitary evolution yields the dephasing map \cite{FMP20}
    \begin{equation}\label{UELTA}
        \overline{\mathcal{U}^\dagger_{SE}(t)}=\sum_a |\Pi_a)(\Pi_a|,
    \end{equation}
    where $\Pi_a=\ket{E_a}\bra{E_a}$ is the energy eigenstate of the total Hamiltonian $H$.
    A unitary evolution $\mathcal{U}^\dagger_{SE}(t)$ together with a dynamical evolution $\mathcal{E}_t$ give the following map
    \begin{align}
        \overline{\mathcal{U}^\dagger_{SE}(t)\circ \dots  \circ \mathcal{E}_t}= \overline{\mathcal{U}^\dagger_{SE}(t)}\circ \dots  \circ \overline{\mathcal{E}_t}\notag \\
        +\sum_{\substack{a,b \\ a\neq b}} |\Pi_{ab})(\Pi_{ab}|\circ \dots  \circ (I_E|\Pi_{ab})(\Pi_{ab}|\rho_E(0)),
    \end{align} 
    where $\Pi_{ab}=\ket{E_b}\bra{E_a}$.
    Correspondingly, the error caused by the second line in \cref{CSMLAVG} can be divided into two parts. The first part is
   \begin{equation}\label{FPETHLAVG}
    ( I_{SE}|[\mathcal{J}_{\rho_{E}(0)}^{1/2}-\mathcal{J}_{\rho^\text{can}_E}^{1/2}] \circ \mathcal{J}_{\rho_S^\text{can}}^{1/2}\circ \overline{\mathcal{N}_\beta^{T'\dagger}(t)}\circ \mathcal{J}^{-1/2}_{\rho^\text{can}_S}|\overline{\rho_S(t)})^\dagger.
   \end{equation}
   We discuss its bound in detail in \cref{EIDESLTR}. The second part is
   \begin{align}\label{GTF2}
       \sum_{\substack{a,b \\ a\neq b}}(\rho_S^\text{can}\otimes[\rho_{E}(0)-\rho^\text{can}_E]|\mathcal{N}^T_{I}(i\beta/2)|\Pi_{ab})^\dagger \times\notag \\
      (\Pi_{ab}| \mathcal{N}^{T\dagger}_{I}(-i\beta /2)\circ\mathcal{J}_{\rho_S^\text{can}}^{-1/2}|\Tr_E \Pi_{ab}\otimes I_E)^\dagger\notag \\
       \times (\Pi_{ab}|\rho_S(0)\otimes\rho_E(0))^\dagger.
   \end{align}
   We discuss its bound in detail in \cref{ECEFS}. 
    Finally, in the large-environment limit, by combining \cref{CSMLAVG,BOGTF1,BOGTF2}, we have
    \begin{align}\label{GFDCL}
        \abs{\delta\overline{ G_F}(i)}\leq  \abs{ \overline{G_F}(i)-\overline{G^{T'}_F}(i)} +\abs{\overline{G^{T'}_F}(i)-\overline{G^{\text{can},T'}_F}(i)} \notag \\
        +\abs{\overline{G^{\text{can},T'}_F}(i)-\overline{G^\text{can}_F}(i)} =o(1),
    \end{align}
    where "$\text{can}$" represents replacing the initial environment state in $ \mathcal{N}_{\text{FT}}(t) $ with the canonical ensemble and "$T'$" denotes the substitution of  $\mathcal{N}_\beta$ in $\mathcal{N}_{\text{FT}}(t) $ with $\mathcal{N}^{T'}_\beta$.
    In conclusion, the first and third terms of \cref{GFDCL} depends on the locality of the imaginary time evolution component of the map $\mathcal{N}_{\beta}(t)$ and can be constrained by the Lieb-Robinson bound. The second term of \cref{GFDCL} depends on the indistinguishability of the states within a typical energy shell and can be restricted by ETH.

    \section{Eigenstate fluctuation theorem for direct heat exchange between two baths}\label{S4}
    We now consider the interaction of two baths $A$ and $B$. The whole system evolves according to the unitary evolution with the Hamiltonian $H=H_A+H_B+H_I$. If the initial state of two baths is the canonical ensemble with different temperatures $\hat{\rho}^\text{can}_X=e^{-\beta_X \hat{H}_X}/Z_X$ and $X=A,B$, then we should choose
    \begin{equation}\label{TCE}
   \hat{\sigma}_t=\beta_A (\hat{H}_A- F_A)+\beta_B (\hat{H}_B- F_B).
    \end{equation}
    If we do not make any assumptions about the initial environment state, but still set $\hat{\sigma}_t$ as \cref{TCE}, then we have
    \begin{align}\label{LMEVTB}
    G_F(\lambda)=(I_{AB}|\mathcal{J}_{\rho^\text{can}_A\otimes\rho^\text{can}_B}^{-i\lambda/2}\circ \mathcal{U}_{AB}(t)\notag \\
    \circ\mathcal{J}_{\rho^\text{can}_A\otimes\rho^\text{can}_B}^{i\lambda/2}\circ  \mathcal{M}_{\hat{\sigma}_0}|\rho_{A}(0)\otimes\rho_{B}(0)).
    \end{align}
    Since the two baths can deviate from the canonical ensemble $\rho^\text{can}$, the integral FT will exhibit deviations. By defining a map like $\mathcal{N}_\beta$ but with different temperatures (see \cref{NBDT}), we can rewrite the conjugated form of \cref{LMEVTB} as $G_F(i)=( I_{AB}| \mathcal{N}_{\text{FT}}(t)|1)^\dagger$, where
    \begin{equation}\label{NFTDFT}
       \mathcal{N}_{\text{FT}}(t):= \mathcal{J}_{\rho'_{A}(0)\otimes\rho'_{B}(0)}^{1/2} \circ   \mathcal{N}^\dagger_{\beta_A,\beta_B}(t).
    \end{equation}
    When the baths $\rho_X(0)$ deviate from the canonical ensemble $\rho^\text{can}_X$, the constructed map $\mathcal{N}_{\text{FT}}(t)$ is generally not a TP map. 

    Here we consider the situation where the initial states of $A$ and $B$ are energy eigenstates. In this case, we have $\rho'_{X}(0)=\rho_{X}(0)$. If the evolution time tends to zero, it is obvious that $ \mathcal{N}_{\text{FT}}(t)= \mathcal{J}_{\rho'_{A}(0)\otimes\rho'_{B}(0)}^{1/2}$. It acts on $|1)$ gives $|\rho'_{A}(0)\otimes\rho'_{B}(0))$.  Therefore, it is also a TP assignment map. For finite time evolution,
   similar to \cref{EFNFT}, we can rewrite the constructed map as
   \begin{align}\label{NPFT}
    \mathcal{N}_{\text{FT}}(t) 
     = \mathcal{J}_{\rho_{A}(0)\otimes\rho_{B}(0)}^{1/2} \circ   [\mathcal{N}^\dagger_{\beta_A,\beta_B}(t)-\mathcal{N}^{T\dagger}_{\beta_A,\beta_B}(t)]\notag\\
     +[\mathcal{J}_{\rho_{A}(0)\otimes\rho_{B}(0)}^{1/2}-\mathcal{J}_{\rho^\text{can}_A\otimes\rho^\text{can}_B}^{1/2} ] \circ\mathcal{N}^{T\dagger}_{\beta_A,\beta_B}(t)\notag \\
     +\mathcal{J}_{\rho^\text{can}_A\otimes\rho^\text{can}_B}^{1/2}\circ [\mathcal{N}^{T\dagger}_{\beta_A,\beta_B}(t)-\mathcal{N}^\dagger_{\beta_A,\beta_B}(t)] \notag\\
     +\mathcal{J}_{\rho^\text{can}_A\otimes\rho^\text{can}_B}^{1/2} \circ   \mathcal{N}^\dagger_{\beta_A,\beta_B}(t) ,
 \end{align}
 where the truncated $\mathcal{N}^{T}_{\beta_A,\beta_B}(t)$ is defined in \cref{NBDT}.  It is a truncated version of $ \mathcal{N}_{\beta_A,\beta_B}(t)$ based on the speed limit of information flow. The last term of \cref{NPFT} is equal to $\mathcal{U}^\dagger_{SE}(t)\circ \mathcal{J}_{\rho^\text{can}_{A}\otimes \rho^\text{can}_{B}}^{1/2}$. It acts on $|1)$ gives $\mathcal{U}^\dagger_{SE}(t)|\rho^\text{can}_{A}\otimes \rho^\text{can}_{B})$. So it's a TP mapping. The terms in the first and third lines of \cref{NPFT}  represent the error due to truncation, which can also be bounded by the Lieb-Robinson bound. We discuss these bounds in detail in \cref{ETSTR}. The term in the second line of \cref{NPFT} represents the error caused by the deviation of the initial state. Since the mapping $\mathcal{N}^{T\dagger}_{\beta_A,\beta_B}(t)$ acts on $|1)$ also gives a local operator, this part of the error can be bounded by ETH. See \cref{EIDES} for details. Finally, in the large-environment limit, by combining the error terms in \cref{NPFT}, we can obtain a similar bound as \cref{EBEFNEF}.  Therefore, the error of FT should vanish in the thermodynamic limit.

 If both heat baths have the same inverse temperature as given by \cref{ITFE}, we can still utilize  the perturbation expansion in \cref{LTETS}. Since the final state measurement here is independent of time, we only need to consider
 \begin{align}\label{TBHELAVG}
    \overline{\mathcal{N}_{\text{FT}}(t)} 
     = \mathcal{J}_{\rho_{A}(0)\otimes\rho_{B}(0)}^{1/2} \circ   \overline{[\mathcal{N}^\dagger_{\beta}(t)-\mathcal{N}^{T'\dagger}_{\beta}(t)]}\notag\\
     +[\mathcal{J}_{\rho_{A}(0)\otimes\rho_{B}(0)}^{1/2}-\mathcal{J}_{\rho^\text{can}_A\otimes\rho^\text{can}_B}^{1/2} ] \circ  \overline{\mathcal{N}^{T'\dagger}_{\beta}(t)}\notag \\
     +\mathcal{J}_{\rho^\text{can}_A\otimes\rho^\text{can}_B}^{1/2}\circ  \overline{[\mathcal{N}^{T'\dagger}_{\beta}(t)-\mathcal{N}^\dagger_{\beta}(t)]} \notag\\
     +\mathcal{J}_{\rho^\text{can}_A\otimes\rho^\text{can}_B}^{1/2} \circ   \overline{ \mathcal{N}^\dagger_{\beta}(t)} ,
 \end{align}
 where $\mathcal{N}_\beta^{T'}(t)$ is defined by \cref{TMLTR}, except that $SE$ is replaced by $AB$ and truncate both $A$ and $B$.
 The last term of \cref{TBHELAVG} is equal to $\overline{\mathcal{U}^\dagger_{AB}(t)}\circ \mathcal{J}_{\rho^\text{can}_A\otimes\rho^\text{can}_B}^{1/2}$, which is also a TP mapping. The terms in the first and third lines of \cref{TBHELAVG} represent the error due to truncation, which can be bounded by the Lieb-Robinson bound. The detailed calculation can be found in \cref{ETSTR}. The term in the second line of \cref{TBHELAVG} represents the error caused by the deviation of the initial state. This part of the error can be bounded with ETH. See \cref{EIDES} for details.  Finally, in the large-environment limit, by combining the error terms in \cref{TBHELAVG}, we can obtain a similar bound as \cref{GFDCL}.  Therefore, the error of FT in the long-time average should vanish in the thermodynamic limit.
 
 If the temperature of the two baths is different, the corresponding term $\rho^\text{can}_A\otimes \rho^\text{can}_B$differs significantly from the imaginary time evolution  $U_{SE}(it)$, rendering the perturbation expansion in \cref{LTETS} inapplicable in this scenario.  Additionally, considering the evolution perspective, since both $A$ and $B$ are large baths, their equilibration time may be considerably long, and they may not reach a steady state quickly. Consequently, it is unlikely that the integral fluctuation theorem holds for a long-time average in this case.

\section{Integral fluctuation theorem from quasi-measurements}\label{S5}
From \cref{TSL}, we observe that, in general, the entropy production in the FTs of open systems is closely linked to the state of the environment. However, when the evolution map possesses a global fixed point, the entropy flux can be expressed solely in terms of system-related quantities \cite{LP21}. Now, let us briefly consider these cases using the method described herein.

For systems with global fixed point
\begin{equation}\label{GFPR}
    \mathcal{U}_{SE}(\rho_S^*\otimes  \rho_E)=\rho^*_S\otimes  \rho_E,
\end{equation}
we have
\begin{equation}\label{RSTFFP}
    \mathcal{J}^{1 /2}_{\rho^{*}_S\otimes  \rho_E}\circ\mathcal{U}_{SE}(t)\circ \mathcal{J}^{-1 /2}_{\rho^{*}_S\otimes  \rho_E}
    =\mathcal{U}_{SE}(t).
\end{equation}
Supposing there is no correlation between the initial system and environment, if the environment is not measured, substituting \cref{RSTFFP} into \cref{GFRSM}, we can get
\begin{align}
    G_F(\lambda)=(I_S\otimes\rho_{E}|\mathcal{J}_{e^{- \hat{\sigma}^S_t}}^{-i\lambda/2}\circ\mathcal{J}_{\rho^{*}_S}^{1/2}\circ\mathcal{U}(t)\notag\\
    \circ\mathcal{J}_{\rho^{*}_S}^{-1/2}\circ\mathcal{J}_{e^{- \hat{\sigma}^S_0}}^{i\lambda/2}\circ \mathcal{M}_{\hat{\sigma}^S_0}|\rho_{S}(0)\otimes I_{E}).
\end{align}
However, it is not possible to choose a suitable $\hat{\sigma}_S(t)$ such that $G_F(i)=1$. The two-point measurement of the local system cannot satisfy the integral FT. Previous work \cite{KK19} established a general Crooks FT for open quantum processes, where only local measurements of the system are required, but these measurements are no longer the conventional two-point measurements. To obtain the so-called quasiprobability, quasi-measurements need to be introduced \cite{H23}. This is a measure of the density matrix and can be reconstructed from positive-operator valued measurements. After incorporating quasi-measurements, the generating function can be expressed as
\begin{align}
    G_F(\lambda):=\sum_{\sigma_0  ,\sigma_t, ij, k'l'} (e^{i\lambda(\hat{\sigma}_t+\hat{\tau}'-\hat{\sigma}_0-\hat{\tau}})| \notag\\
    \Pi_{\sigma_t}\otimes \Pi_{\sigma_0}\otimes \Pi_{ij}\otimes \Pi_{k'l'})(O^{(\sigma_t,\sigma_0,ij,k'l')}|\mathcal{S}),
\end{align}
where $\{\ket{i}\}$ is the diagonal basis of $\hat{\tau}$ and $\{\ket{k'}\}$ is the diagonal basis of $\hat{\tau}'$. The process state $\mathcal{S}$ is the Choi-state form of the process tensors and $O^{(\sigma_t,\sigma_0,ij,k'l')}$ is the Choi-state of the measurement. For specific definitions, refer to \cite{H23}.
Unlike \cref{EXP2RSM}, according to the completeness relation, we have
\begin{equation}
    \sum_{ij}  (e^{-i\lambda\hat{\tau}}| \Pi_{ij})| \Pi_{ij})( \Pi_{ij}| =\mathcal{J}_{e^{- \hat{\tau}}}^{i\lambda/2}.
\end{equation}
Similar to the procedure used in \cref{GFRSM}, we can rewrite the generating function as 
\begin{align}\label{QUASIGF}
    G_F(\lambda)=(I_S|\mathcal{J}_{e^{- \hat{\sigma}^S_t}}^{-i\lambda/2}\circ\mathcal{J}_{e^{-\tau'}}^{-i\lambda/2}\circ\mathcal{N}(t)\notag\\
    \circ\mathcal{J}_{e^{-\tau}}^{i\lambda/2}\circ\mathcal{J}_{e^{- \hat{\sigma}^S_0}}^{i\lambda/2}\circ \mathcal{M}_{\hat{\sigma}^S_0}|\rho_{S}(0)),
\end{align}
where $\mathcal{N}(t)=(I^E|\mathcal{U}(t)|\rho_E)$. We can set $\hat{\sigma}_t=-\ln \hat{\rho}_S(t)$ and $\tau=\tau'=\ln \rho^{*}_S$, then the conjugated form of generating function gives
\begin{align}\label{NEORC}
    G_F(i-\lambda)=(I_{S}|\mathcal{J}_{e^{- \hat{\sigma}^S_0}}^{-i\lambda/2} \circ\mathcal{J}_{e^{-\tau}}^{-i\lambda/2} \circ\mathcal{J}_{\rho^{*}_S}^{1/2}\circ\mathcal{N}^\dagger(t) \notag\\
   \circ\mathcal{J}_{\rho^{*}_S}^{-1/2}\circ\mathcal{J}_{e^{-\tau'}}^{i\lambda/2}\circ \mathcal{J}_{e^{- \hat{\sigma}^S_t}}^{i\lambda/2}|\rho_S(t))^\dagger.
\end{align}
From this formula, we can naturally obtain a backward process map
\begin{equation}
    \mathcal{R}(t)=\mathcal{J}_{\rho^{*}_S}^{1/2}\circ\mathcal{N}^\dagger(t)  \circ\mathcal{J}_{\mathcal{N}(\rho^{*}_S)}^{-1/2},
\end{equation}
which is just the Petz recovery map used in \cite{MHP15,MHP18,KK19}. From \cref{NEORC}, we know that the integral FT is guaranteed by the TP property of the Petz recovery map
\begin{equation}
    G_F(i)=(I|\mathcal{R}(t)|\rho_S(t))^\dagger=1.
\end{equation}
Different from \cref{NCSE}, the entropy production here depends only on the local state of the system and does not need to know the state of the environment.

\section{Conclusion and discussion}\label{S6}
In this paper, we have redefined the generating function of the general FT under unitary evolution and obtained a constructed map. The initial state, measurements, their Fourier transform, and evolution are all encompassed in this map. The constructed map is CP, and its TP property determines the integral FT. This unified form can help simplify the impact of initial state deviation on the FT. In particular, for the measurement of energy $\hat{H}$, the corresponding rescaling map is imaginary time evolution. The commutativity between imaginary time evolution and real-time evolution can effectively simplify calculations. With the help of this new formalism, we have proved the eigenstate FT in a simpler manner that requires fewer and weaker assumptions. The rationale behind these proofs is roughly as follows: The error of the FT can be divided into two components. One arises from the non-locality of the map $\mathcal{N}_{\beta}(t)$ and can be bounded by the Lieb-Robinson bound. The other originates from the distinguishability of environment states and can be restricted by ETH. Both errors vanish in the thermodynamic limit, ensuring the FT's validity even when the initial state of the bath is a single energy eigenstate of a many-body systems.

We also explored the heat exchange between two baths. The initial state of both baths is a pure state. We observed that the integral FT remains approximately valid in the short-time regime. In the long-time regime, when the temperatures of the two baths are equal, the long-time average of the integral FT holds approximately. However, when the temperatures differ, the long-time average of the integral FT may not hold.

Referring to the FTs for the quantum channel, we have examined the generating function of the quasi-probability. The Petz recovery map can be naturally derived from this generating function. The integral FT is ensured by the TP property of the Petz recovery map.

An additional advantage of our method is its ease of generalization to multipoint measurements. The operator-state formalism employed here seamlessly integrates with process tensors \cite{H23}. Process tensors offer the convenience of studying multitime processes and multipoint measurements \cite{PT,PTR2}. FTs beyond two-point measurements have garnered extensive attention \cite{HV10,LRJ12,CS18,MLL20,JPW14,BB20}. The integral FT under multitime processes can aid in exploring the generalization of the fluctuation-dissipation theorem for n-point systems \cite{WH02}. Since the memory effect can result in a negative entropy production rate \cite{H23}, research in this area also contributes to a deeper understanding of the second law of thermodynamics: The second law implies complete irreversibility, while the Poincar\'e recurrence theorem corresponds to complete reversibility. FTs enable us to derive irreversibility from reversible evolution, while the memory effect can weaken irreversibility and yield negative entropy production rates. Investigating the influence of the ETH on FTs under multitime evolution allows us to further comprehend the impact and limitations of the memory effect. This, in turn, aids in understanding the role of the thermodynamic limit in the second law of thermodynamics and the Poincar\'e recurrence theorem.

\begin{acknowledgments}
    Z.H. is supported by the National Natural Science Foundation of
China under Grants No. 12305035. We also thank X.-K. Guo for helpful comments.
\end{acknowledgments}

\appendix
\section{The crucial lemmas and hypothesis used from the literature}
The lemma 3.1 in \cite{AKL16} tells us that when $0\leq s<\frac{1}{gk}$, we have
\begin{equation}
    \norm{e^{s H }A e^{-s H}}\leq \norm{A }\times (1-gks)^{-R/gk},
\end{equation}
where $H$ is a local Hamiltonian, and $A$ is an arbitrary local operator. The parameters $g$, $k$, and $R$ are related to the lattice structure and interaction between particles and are independent of the system size. According to this lemma, as long as the system's Hamiltonian meets the corresponding requirements, when $\tau$ is sufficiently small, the term in \cref{ITTE} can be bounded as
\begin{equation}\label{HIGRB}
    \norm{H_I(i\tau)}=\Theta(1)\times\norm{H_I}.
\end{equation}

 The lemma 5 in \cite{HHKL21} gives the following Lieb-Robinson bound
 \begin{equation}
    \norm{e^{i t H} O_X e^{-i t H}-e^{i t H^T} O_X e^{-i t H^T}}\leq \abs{X}\norm{O_X}\frac{(2\zeta_0\abs{t})^{\ell}}{\ell!},
 \end{equation}
 where $H$ is a local Hamiltonian, $H^T$ is the truncated Hamiltonian, and $O_X$ is any local operator. The parameter $\ell$ represents the distance between $O_X$ and the truncation boundary, and $\abs{X}$ denotes the number of sites in the support $X$. The constant $\zeta_0=O(1)$ depends on the lattice structure and interactions among lattice sites. According to this lemma, the error induced by truncation will be exponentially small
\begin{equation}\label{LRB}
    \norm{\delta H_I(\tau)}=\abs{H_I}\times \norm{H_I}\times O(e^{-\mu \ell})
\end{equation}
as long as the evolution time $\abs{\tau}\leq \ell/v_{LR}$, where $v_{LR}$ is so-called Lieb-Robinson velocity.

According to the canonical typicality \cite{DLL18}, weak ETH \cite{IKS17,KS20} or subsystem ETH \cite{HG23}, we know that it is difficult for local operators to distinguish the global energy eigenstates from the canonical ensembles
\begin{align}\label{DGETH}
   \abs{ (\Pi_{a}|O)-(\rho^\text{can}|O)}\leq \norm{O}\times N^{-\alpha}, \\ \label{OFFDGETH}
   \abs{ (\Pi_{ab}|O)}\leq \norm{O}\times N^{-\alpha} \quad (a\neq b), 
\end{align}
The parameter $0<\alpha$. $N$ denotes the size (the number of sites) of the entire system. The size of the local operator $O$ should be much smaller than $N$. It is worth noting that in \cref{DGETH}, the errors decay algebraically, which is weaker than the exponential decay usually expected in ETH. Therefore, it can be anticipated that this hypothesis can be satisfied in a broader range of systems.

\section{The properties of $\mathcal{N}_\beta$}\label{PPMB}
\subsection{Imaginary time evolution and strict energy conservation condition}\label{ITESECC}
The superoperator $\mathcal{N}_\beta$ of \cref{SOFNB} can also be written as
\begin{align}\label{NBWST}
    \mathcal{N}_\beta(t)(\cdot)=\mathcal{U}_{SE}^0(-\beta \lambda/2)\circ\mathcal{U}_{SE}(t)\circ \mathcal{U}_{SE}^0(\beta \lambda/2)(\cdot) \notag \\
    =\mathcal{M}_{\lambda}(t)(\cdot)\mathcal{M}_{\lambda}^{\dagger}(t),
 \end{align}
 where 
 \begin{equation}\label{MLF}
     \mathcal{M}_{\lambda}(t)=U^0_{SE}(-\beta \lambda/2)[U_{SE}(t)]U^0_{SE}(\beta \lambda/2)
 \end{equation}
 and 
 \begin{equation}
     U^0_{SE}(\tau)=\exp(-i(H_S+H_E)\tau).
 \end{equation}
 When $\lambda$ is a pure imaginary number, the mapping $\mathcal{U}_{SE}^0(-\beta \lambda/2)$ is just the imaginary time evolution \cite{VGC04}. Unlike the real time evolution, imaginary time evolution is CP but not TP. From \cref{DFUSPO,CJRSPO}, we have
 \begin{equation}
     \mathcal{U}^{0\dagger}_{SE}(\tau)= \mathcal{U}^{0}_{SE}(-\tau^*).
 \end{equation}
 Therefore, \emph{the imaginary time evolution is self-conjugate}. Since $U_{SE}(t)=e^{-i(H_0+H_I)t}$ and $ [U^0_{SE},H_0]=0$, we can rewrite \cref{MLF} as
 \begin{equation}\label{MLF1}
     \mathcal{M}_{\lambda}(t)=e^{-i[H_0+H_I(\beta \lambda/2)]t},
 \end{equation}
 where
 \begin{equation}\label{ITTE}
     H_I(\tau):=U^0_{SE}(-\tau) H_I U^0_{SE}(\tau).
 \end{equation}
 Notice that it is different from the Hermitian operator
 \begin{equation}
     \mathcal{U}^0_{SE}(-\tau) H_I=U^0_{SE}(-\tau) H_I U^{0\dagger}_{SE}(-\tau).
 \end{equation}
 $H_I(\beta \lambda/2)$ is a pseudo-Hermitian operator \cite{BF09}. It is easy to verify that with $\rho =e^{2 H_0 \text{Im}(\tau)}$, we have $\rho H_I(\tau) \rho^{-1}=H^\dag_I(\tau)$. According to \cref{MLF1}, if there is a strict energy conservation condition $[H_0,H_I]=0$ \cite{DMK22}, or the temperature tends to infinity, there will be $\mathcal{M}_{\lambda}(t)=U_{SE}(t)$. And then we have $\mathcal{N}_\beta(t)=\mathcal{U}_{SE}(t)$.
\subsection{Short-time evolution and truncated superoperator}\label{STETS}
 The evolution map can be written in the integral form 
    \begin{align}\label{EHI}
        \mathcal{U}_{SE}(t)=\mathcal{U}_{SE}(t)\circ\mathcal{U}^0_{SE}(0)\notag\\
        =\mathcal{U}^0_{SE}(t)+\int_0^t d\tau \frac{\partial}{\partial_\tau} \mathcal{U}_{SE}(\tau )\circ\mathcal{U}^0_{SE}(t-\tau)\notag\\
        =\mathcal{U}^0_{SE}(t)+\int_0^t d\tau \mathcal{U}_{SE}(\tau )\circ \mathcal{L}_I\circ\mathcal{U}^0_{SE}(t-\tau)
    \end{align}
    where $\mathcal{L}_I (\cdot):=-i[H_I,\cdot]$. Here we only take the first-order perturbation approximation
    \begin{equation}\label{UFOPA}
        \mathcal{U}_{SE}(t)\simeq\mathcal{U}^0_{SE}(t)[\mathcal{I}+ \int_0^t d\tau  \mathcal{U}^0_{SE}(\tau-t)\circ \mathcal{L}_I\circ \mathcal{U}^0_{SE}(t-\tau)].
    \end{equation}
    The error caused by taking the first-order approximation 
    \begin{equation}
       \delta  \mathcal{U}_{SE}(t)= \int_0^t d\tau [\mathcal{U}_{SE}(\tau )-\mathcal{U}^0_{SE}(\tau)]\circ \mathcal{L}_I\circ\mathcal{U}^0_{SE}(t-\tau).
    \end{equation}
    Using integral form \cref{EHI}, the error can be rewritten as
    \begin{equation}\label{FOAPER}
        \int_0^t d\tau_1\int_0^{\tau_1} d\tau_2 \mathcal{U}_{SE}(\tau_2 )\circ \mathcal{L}_I\circ\mathcal{U}^0_{SE}(\tau_1-\tau_2) \circ \mathcal{L}_I\circ\mathcal{U}^0_{SE}(t-\tau_1).
    \end{equation}
    After it acts on arbitrary operator $O$, the bound of error can be calculated with
    \begin{align}
        \norm{\delta  \mathcal{U}_{SE}(t) O}\leq \int_0^t d\tau_1\int_0^{\tau_1} d\tau_2 \lVert \mathcal{U}_{SE}(\tau_2 )\circ\mathcal{U}^0_{SE}(-\tau_2 )\notag\\
        \circ\mathcal{L}_I(-\tau_2)\circ  \mathcal{L}_I(-\tau_1)\circ\mathcal{U}^0_{SE}(t)O \rVert\leq \frac{(2t\norm{H_I})^2}{2}\norm{O},
    \end{align}
    where $ \mathcal{L}_I(\tau) O=-i \left[ H_I(\tau)O-O H^\dag_I(\tau)\right]$.
    The error is small when the evolution time is short enough $t<<1/(2\norm{H_I})$.
    Using the approximation \cref{UFOPA}, it is easy to show that
    \begin{widetext}
    \begin{align}\label{1HIAP}
        \mathcal{N}^\dagger_{\beta}(t)(O_S\otimes I_E)= \mathcal{J}^{i\lambda /2}_{\rho^\text{can}_S\otimes \rho^\text{can}_E}\circ\mathcal{U}^\dagger_{SE}(t)\circ \mathcal{J}^{-i\lambda /2}_{\rho^\text{can}_S\otimes \rho^\text{can}_E}(O_S\otimes I_E)\notag \\
        \simeq \mathcal{U}^{0\dagger}_{SE}(t)(O_S\otimes I_E)+\int_0^t d\tau  \mathcal{U}^{0\dagger}_{SE}(t-\tau)\circ \mathcal{J}^{i\lambda /2}_{\rho^\text{can}_S\otimes \rho^\text{can}_E}\circ \mathcal{L}^\dagger_I\circ \mathcal{J}^{-i\lambda /2}_{\rho^\text{can}_S\otimes \rho^\text{can}_E}\circ \mathcal{U}^{0\dagger}_{SE}(\tau-t) (\mathcal{U}^{0\dagger}_{SE}(t)O_S\otimes I_E)\notag \\
        =\mathcal{U}^{0\dagger}_{S}(t)(O_S\otimes I_E)+\int_0^t d\tau  \mathcal{U}^{0\dagger}_{SE}(t-\tau)\circ\mathcal{U}_{SE}^{0\dagger}(-\beta \lambda/2)\circ \mathcal{L}^\dagger_I\circ\mathcal{U}_{SE}^{0\dagger}(\beta \lambda/2)\circ  \mathcal{U}^{0\dagger}_{SE}(\tau-t) (\mathcal{U}^{0\dagger}_{S}(t)O_S\otimes I_E)\notag\\
        =\mathcal{U}^{0\dagger}_{S}(t)(O_S\otimes I_E)+\int_0^t d\tau \mathcal{L}^\dagger_I(t -\tau-\beta \lambda/2) (\mathcal{U}^{0\dagger}_{S}(t)O_S\otimes I_E)
    \end{align}
    where $ \mathcal{L}^\dagger_I(\tau) O=i \left[ H_I(\tau)O-O H^\dag_I(\tau)\right]$. The first term on the right side of the last equation of (\ref{1HIAP}) only contains the local evolution of the system, and its contribution to $G_F(\lambda)$ is
    \begin{equation}
    \Tr[\mathcal{J}_{\rho_S(t)}^{-i\lambda/2}\circ\mathcal{J}_{\rho_S^\text{can}}^{i\lambda/2}\circ\mathcal{U}^0_{S}(t)\circ\mathcal{J}_{\rho_S^\text{can}}^{-i\lambda/2}\circ\mathcal{J}_{\rho_S(0)}^{i\lambda/2}(\rho_{S}(0)\otimes\rho'_{E}(0))]=\Tr[\mathcal{J}_{\rho_S(t)}^{-i\lambda/2}\circ\mathcal{U}^0_{S}(t)\circ\mathcal{J}_{\rho_S(0)}^{i\lambda/2}(\rho_{S}(0))]\overset{\lambda=i}{=}1.
    \end{equation}
   Now, let's analyze the contribution of the second term. We divide the environment into two parts: $B_0$, which is the portion close to the system, and $\overline{B_0}$, which comprises the remaining parts. When the temperature is high and the evolution time is short, the scale of $B_0$ can be larger than the propagation range of the Lieb-Robinson velocity $v_{LR}(\abs{t+\beta \lambda/2})$, but much smaller than the overall environment scale $L_{E}$. The Hamiltonian of the environment can also be divided as $H_E=H_{B_0}+H_{\overline{B_0}}+H_{B_0 \overline{B_0}}$. We define the truncated Hamiltonian $H^T_{0}:=H_S+H_{B_0}$, the corresponding time evolution operator $U^T_{SE}=e^{-i(H_{0,T}+H_I)t}$, the truncated canonical ensemble $\rho^\text{can}_{B_0}=e^{-\beta H_{B_0}}/Z_{B_0}$, and the corresponding superoperator $\mathcal{N}^{T}_{\beta}(t):=\mathcal{J}^{-i\lambda /2}_{\rho^\text{can}_S\otimes \rho^\text{can}_{B_0}}\circ\mathcal{U}^T_{SE}(t)\circ \mathcal{J}^{i\lambda /2}_{\rho^\text{can}_S\otimes \rho^\text{can}_{B_0}}$. Following a similar procedure to that in \cref{EHI,1HIAP}, we have:
\begin{equation}\label{TREDNB}
    \mathcal{N}^{T\dagger}_{\beta}(t)(O_S\otimes I_E)\simeq\mathcal{U}^{0\dagger}_{S}(t)(O_S\otimes I_E)+\int_0^t d\tau \mathcal{L}^{T\dagger}_I(t -\tau-\beta \lambda/2) (\mathcal{U}^{0\dagger}_{S}(t)O_S\otimes I_E),
\end{equation}
where $ \mathcal{L}^{T\dagger}_I(\tau )O=i \left[ H^T_I(\tau )O-O H^{T\dag}_I(\tau )\right]$. $H^T_I(\tau):=e^{iH^T_0\tau } H_I e^{-iH^T_0\tau }$ is also a pseudo-Hermitian operator. 

\subsection{Long-time evolution and truncated superoperator}\label{LTETS}
When the evolution time is short, employing perturbation theory for the time evolution $\mathcal{U}_{SE}(t)$ is suitable. However, this method may not be appropriate for long evolution times. An alternative approach, as demonstrated in \cite{IKS22}, involves utilizing perturbation theory for the eigenstates of $H$. Here, we explore a different strategy. Notably, when the Hamiltonian remains constant, the imaginary-time evolution and the real-time evolution commute. Therefore, the superoperator $\mathcal{N}_\beta(t)$ can be expressed as
\begin{equation}\label{CMRORIU}
    \mathcal{N}_\beta(t)= \mathcal{N}_{I}(-\beta \lambda/2)\circ\mathcal{U}_{SE}(t)\circ\mathcal{N}^\dagger_{I}(\beta \lambda/2),
\end{equation}
where $  \mathcal{N}_{I}(-\beta \lambda/2):=\mathcal{U}_{SE}^{0}(-\beta\lambda/2)\circ \mathcal{U}_{SE}(\beta\lambda/2)$.
The imaginary time evolution can also be written in the integral form
    \begin{equation}
        \mathcal{U}_{SE}(it)=\mathcal{U}^0_{SE}(0)\circ\mathcal{U}_{SE}(it)
        =\mathcal{U}^0_{SE}(it)+\int_0^t d\tau \frac{\partial}{\partial_\tau} \mathcal{U}^0_{SE}(it-i\tau)\circ\mathcal{U}_{SE}(i\tau )
        =\mathcal{U}^0_{SE}(it)\left[\mathcal{I}+\int_0^t d\tau \mathcal{U}^0_{SE}(-i\tau)\circ \mathcal{L}^{\text{A}}_I\circ\mathcal{U}_{SE}(i\tau )\right],
    \end{equation}
    where $\mathcal{L}^{\text{A}}_I (\cdot):=\{H_I,\cdot\}$. Similar to \cref{FOAPER}, the error caused by taking the first-order approximation 
    \begin{equation}
        \delta  \mathcal{U}_{SE}(it)= \int_0^t d\tau_1\int_0^{\tau_1} d\tau_2\mathcal{U}^0_{SE}(it-i\tau_1)\circ \mathcal{L}^A_I\circ\mathcal{U}^0_{SE}(i\tau_1-i\tau_2)\circ \mathcal{L}^A_I \circ \mathcal{U}_{SE}(i\tau_2 ).
    \end{equation}
    After it acts an arbitrary operator $O$, the bound of error can be calculated with
    \begin{align}
        \norm{\mathcal{U}^0_{SE}(-it)\circ\delta  \mathcal{U}_{SE}(it) O}\leq \int_0^t d\tau_1\int_0^{\tau_1} d\tau_2 \lVert \mathcal{L}^A_I(i\tau_1)\circ  \mathcal{L}^A_I(i\tau_2) \circ  \mathcal{U}^0_{SE}(-i\tau_2 )\circ \mathcal{U}_{SE}(i\tau_2 )O \rVert\notag \\
        \leq \frac{(2t\norm{H_I(i\tau_{\max})})^2}{2}\norm{\mathcal{N}_{I}(-i\tau'_{\max})O}
    \end{align}
    where $\mathcal{L}^{\text{A}}_I(i\tau) O=H_I(i\tau)O+OH^\dagger_I(i\tau)$. The $\tau_{\max}\in [0,\beta/2]$ makes $\norm{H_I(i\tau_{\max})}$ maximize and  $\tau'_{\max}\in [0,\beta/2]$ makes $\norm{\mathcal{N}_{I}(-i\tau'_{\max})O}$  maximize. According to \cref{HIGRB}, the error is small when the temperature is high enough $\beta<<\frac{1}{\Theta(1)\norm{H_I}}$.
    Applying the first-order perturbation approximation to the imaginary time evolution, we obtain
    \begin{equation}\label{ITEFOP}
        \mathcal{N}_{I}(-it)\simeq \mathcal{I}+\int_0^t d\tau \mathcal{L}^{\text{A}}_I(i\tau ).
    \end{equation}
    Similar to the procedure in \cref{STETS}, we can define the truncated superoperator $\mathcal{U}_{SE}^{0,T}$ and $\mathcal{N}^T_{I}(-\beta \lambda/2)$.  Employing a similar approximation as in (\ref{ITEFOP}), we have
    \begin{equation}\label{NTAIE}
        \mathcal{N}^{T}_{I}(-it) :=\mathcal{U}_{SE}^{0,T}(-it)\circ \mathcal{U}^T_{SE}(it) \simeq \mathcal{I}+\int_0^t d\tau  \mathcal{L}^{\text{A},T}_I(i\tau ).
    \end{equation}
    Accordingly, we can define a truncated map
    \begin{align}\label{TMLTR}
        \mathcal{N}^{T'}_{\beta}(t):= \mathcal{N}^T_{I}(-\beta \lambda/2)\circ\mathcal{U}_{SE}(t)\circ\mathcal{N}^{T\dagger}_{I}(\beta \lambda/2).
    \end{align}

    \subsection{$\mathcal{N}_\beta$ with different temperature}\label{NBDT}
   In \cref{NBWST}, we assume that both parts have the same temperature. However, when the temperatures of the two parts differ, the expression for $\mathcal{J}_{\rho^\text{can}_A\otimes \rho^\text{can}_B}$ in terms of $\mathcal{U}_{AB}^0$ is no longer applicable. Instead, we can use the following mapping
    \begin{equation}
        \mathcal{N}_{\beta_A,\beta_B}(t)(\cdot):= \mathcal{J}^{-i\lambda /2}_{\rho^\text{can}_A\otimes \rho^\text{can}_B}\circ\mathcal{U}_{AB}(t)\circ \mathcal{J}^{i\lambda /2}_{\rho^\text{can}_A\otimes \rho^\text{can}_B}(\cdot)=\mathcal{U}_{AB}^0(\beta_A,\beta_B,-\lambda/2)\circ\mathcal{U}_{AB}(t)\circ \mathcal{U}_{AB}^0(\beta_A,\beta_B,\lambda/2) 
        \end{equation}
    where $U^0_{AB}(\beta_A,\beta_B,\tau):=\exp(-i(H_A \beta_A\tau+H_E\beta_B\tau))$.
     Since the temperature of the two systems is different, the duration of imaginary time evolution time is also different. We can define
    \begin{equation}
        H_I(\tau_A,\tau_B):=U^0_{A}(-\tau_A) U^0_{B}(-\tau_B)H_I U^0_{A}(\tau_A) U^0_{B}(\tau_B).
    \end{equation}
    Similar to \cref{1HIAP}, it is easy to prove
    \begin{equation}
        \mathcal{N}^\dagger_{\beta_A,\beta_B}(t)\simeq 
        \mathcal{I} +\int_0^t d\tau \mathcal{L}^\dagger_I(t -\tau-\beta_A \lambda/2,t -\tau-\beta_B \lambda/2) ,
    \end{equation}
    where $\mathcal{L}^\dagger_I(\tau_A,\tau_B) O=i \left[ H_I(\tau_A,\tau_B)O-O H^\dag_I(\tau_A,\tau_B)\right]$. We can partition $X$ (A and B) into two parts: $X_0$, the part close to the sites of $H_I$, and $X_1$, the part far away from $H_I$. When the temperature is high and the evolution time is short, the scale of $X_0$ can exceed the propagation range of the Lieb-Robinson velocity $v_{LR}(\abs{t+\beta_X \lambda/2})$, but remains much smaller than the overall scale $L_{X}$. The Hamiltonian of the environment can also be divided as $H_X=H_{X_0}+H_{X_1}+H_{X_0 X_1}$. We define the truncated Hamiltonian $H^T_{0}:=H_{A_0}+H_{B_0}$, the corresponding time evolution operator $U^T_{AB}=e^{-i(H_{0,T}+H_I)t}$, and the corresponding superoperator $\mathcal{N}^{T}_{\beta}(t):=\mathcal{J}^{-i\lambda /2}_{\rho^\text{can}_{A_0}\otimes \rho^\text{can}_{B_0}}\circ\mathcal{U}^T_{AB}(t)\circ \mathcal{J}^{i\lambda /2}_{\rho^\text{can}_{A_0}\otimes \rho^\text{can}_{B_0}}$. Similar to \cref{TREDNB}, we have:
    \begin{equation}\label{NTDFT}
        \mathcal{N}^{T\dagger}_{\beta_A,\beta_B}(t)\simeq 
        \mathcal{I} +\int_0^t d\tau \mathcal{L}^{T\dagger}_I(t -\tau-\beta_A \lambda/2,t -\tau-\beta_B \lambda/2) .
    \end{equation}
\section{Error estimation}\label{EESEC}
\subsection{The error induced by truncation}\label{ETSTR}
According to \cref{IFTDT,TREDNB}, the difference between $G_{F}(i)$ and the truncated one is
\begin{align}
    \delta G_{LR}(i):= G_{F}(i)- G^T_{F}(i)= ( I_{SE}| \mathcal{J}_{\rho_{E}(0)}^{1/2} \circ \mathcal{J}_{\rho_S^\text{can}}^{1/2}\circ [\mathcal{N}_\beta^\dagger(t)-\mathcal{N}_\beta^{T\dagger}(t)]\circ \mathcal{J}^{-1/2}_{\rho^\text{can}_S}|\rho_S(t))^\dagger\notag \\
    \simeq \int_0^t d\tau( I_{SE}| \mathcal{J}_{\rho_{E}(0)}^{1/2} \circ \mathcal{J}_{\rho_S^\text{can}}^{1/2}\circ(\mathcal{L}^{\dagger}_I(t-\tau -i\beta /2)-\mathcal{L}^{T\dagger}_I(t -\tau-i\beta /2))  \circ\mathcal{U}^{0\dagger}_{S}(t)\circ\mathcal{J}_{\rho_S^\text{can}}^{-1/2}|\rho_S(t))^\dagger\notag\\
    =\int_0^t d\tau( I_{SE}| \mathcal{J}_{\rho_{E}(0)}^{1/2} \circ\mathcal{U}^{0}_{S}(t-\tau-i\beta/2)\circ[\mathcal{L}^{\dagger}_I(t-\tau -i\beta /2)-\mathcal{L}^{T\dagger}_I(t -\tau-i\beta /2)]\circ\mathcal{U}^{0}_{S}(\tau-t+i\beta/2)\circ\mathcal{U}^{0\dagger}_{S}(\tau)|\rho_S(t))^\dagger.
\end{align}
Using this expression, we can constrain the difference 
\begin{align}\label{DGLR}
    \abs{\delta G_{LR}(i) } \simeq \abs{i\int_0^t d\tau \Tr \left\{\rho_{E}(0)e^{-i H_S(t-\tau -i\beta /2 )}\delta H_I(t-\tau -i\beta /2)e^{i H_S(t-\tau -i\beta /2 )}[\mathcal{U}^{0\dagger}_{S}(\tau)\rho_S(t)]\right\}+\text{h.c.}}\notag\\
    \leq 2\int_0^t d\tau \norm{e^{-i H_S(t-\tau -i\beta /2 )}\delta H_I(t-\tau -i\beta /2)e^{i H_S(t-\tau -i\beta /2 )}}\times\norm{\mathcal{U}^{0\dagger}_{S}(\tau)\rho_S(t)\otimes  \rho_{E}(0)}\leq 2\int_0^t d\tau \norm{\delta H'_I(t-\tau -i\beta /2)},
\end{align}
where $\delta H'_I(\tau)=e^{iH_E\tau } H_I e^{-iH_E\tau }-e^{iH_{B_0}\tau } H_I e^{-iH_{B_0}\tau }$. In the second line, we have utilized the Cauchy-Schwartz inequality.  According to \cref{LRB}, we have $\abs{\delta G_{LR}(i) }=o(1)$ if $\abs{t+i\beta/2}<  \ell/v_{LR}$. It's important to note that we employ the first-order perturbation approximation in this calculation and throughout the remainder of this section.

According to \cref{IFTDT,NTAIE}, the difference $ \delta G'_{LR}(i):= G_{F}(i)- G^{T'}_{F}(i)$ is equal to
\begin{align}
   ( I| \mathcal{J}_{\rho_{E}(0)}^{1/2} \circ \mathcal{J}_{\rho_S^\text{can}}^{1/2}\circ [\mathcal{N}_{I}(i\beta/2)\circ\mathcal{U}^\dagger_{SE}(t)\circ\mathcal{N}^{\dagger}_{I}(-i\beta /2)-\mathcal{N}^{T}_{I}(i\beta /2)\circ\mathcal{U}^\dagger_{SE}(t)\circ\mathcal{N}^{T\dagger}_{I}(-i\beta /2)]\circ \mathcal{J}^{-1/2}_{\rho^\text{can}_S}|\rho_S(t))^\dagger\notag \\
    \simeq \int_0^{-\beta/2} d\tau_1( I| \mathcal{J}_{\rho_{E}(0)}^{1/2} \circ \mathcal{J}_{\rho_S^\text{can}}^{1/2}\circ\left[\mathcal{L}^{\text{A}}_I(i\tau_1 )- \mathcal{L}^{\text{A},T}_I(i\tau_1 )\right]  \circ\mathcal{U}^\dagger_{SE}(t)\circ[I+\int_0^{\beta/2} d\tau_2 \mathcal{L}^{\text{A}\dagger}_I(i\tau_2 )]\circ\mathcal{J}_{\rho_S^\text{can}}^{-1/2}|\rho_S(t))^\dagger\notag\\
    +\int_0^{\beta/2} d\tau_2( I| \mathcal{J}_{\rho_{E}(0)}^{1/2} \circ \mathcal{J}_{\rho_S^\text{can}}^{1/2}\circ [I+\int_0^{-\beta/2} d\tau_1\mathcal{L}^{\text{A},T}_I(i\tau_1 )] \circ\mathcal{U}^\dagger_{SE}(t)\circ\left[ \mathcal{L}^{\text{A}\dagger}_I(i\tau_2 )- \mathcal{L}^{\text{A},T\dagger}_I(i\tau_2 )\right]\circ\mathcal{J}_{\rho_S^\text{can}}^{-1/2}|\rho_S(t))^\dagger.
\end{align}
Using this expression, we can constrain the difference 
\begin{align}
    \abs{\delta G'_{LR}(i)}\simeq \abs{\int_0^{-\beta/2} d\tau_1 \Tr\left\{\rho_S^\text{can}\otimes  \rho_{E}(0)\delta H_I(i\tau_1)[\mathcal{U}^\dagger_{SE}(t)\circ(I+\int_0^{\beta/2} d\tau_2 \mathcal{L}^{\text{A}\dagger}_I(i\tau_2 ))\circ\mathcal{J}_{\rho_S^\text{can}}^{-1/2}\rho_S(t)]\right\}+\text{h.c.}}\notag\\
    +\abs{\int_0^{\beta/2} d\tau_2 \Tr\left\{[\mathcal{U}_{SE}(t)\circ(I+\int_0^{-\beta/2} d\tau_1\mathcal{L}^{\text{A},T}_I(i\tau_1 ))\rho_S^\text{can}\otimes  \rho_{E}(0)]\delta H_I(i\tau_2)[\mathcal{J}_{\rho_S^\text{can}}^{-1/2}\rho_S(t)]\right\}+\text{h.c.}}\notag\\
    \leq 2\int_0^{\beta/2} d\tau \norm{\rho_S^\text{can}\otimes  \rho_{E}(0)}\times \norm{\delta H_I(i\tau)}\times \left[2\norm{I}+2\int_0^{\beta/2} d\tau'(\norm{H_I(i\tau')}+\norm{H^T_I(i\tau')})\right]\times \norm{\mathcal{J}_{\rho_S^\text{can}}^{-1/2}\rho_S(t)}.
\end{align}
The term $\norm{\mathcal{J}_{\rho_S^\text{can}}^{-1/2}\rho_S(t)}$  is related to the final state of the system. It can be greater than $1$ but will not increase with the size of the environment.
According to \cref{HIGRB,LRB}, we have $\abs{\delta G'_{LR}(i) }=o(1)$ if $\abs{i\beta/2}<  \ell/v_{LR}$. Since $\abs{\overline{\delta G'_{LR}(i)} }\leq \overline{\abs{\delta G'_{LR}(i)} }=o(1)$,  the error is very small under long-time average.

According to \cref{NFTDFT,NTDFT}, the error induced by truncation in \cref{NPFT} can be estimated as
\begin{align}
    \delta G_{LR}(i)= ( I_{AB}|  \mathcal{J}_{\rho_{A}(0)\otimes\rho_{B}(0)}^{1/2} \circ   [\mathcal{N}^\dagger_{\beta_A,\beta_B}(t)-\mathcal{N}^{T\dagger}_{\beta_A,\beta_B}(t)]|1)^\dagger\notag \\
    \simeq \int_0^t d\tau( I_{AB}|  \mathcal{J}_{\rho_{A}(0)\otimes\rho_{B}(0)}^{1/2} \circ   [\mathcal{L}^{\dagger}_I(t -\tau-\beta_A \lambda/2,t -\tau-\beta_B \lambda/2)-\mathcal{L}^{T\dagger}_I(t -\tau-\beta_A \lambda/2,t -\tau-\beta_B \lambda/2)]|1)^\dagger
\end{align}
Using this expression, we can constrain the difference as
\begin{align}
    \abs{\delta G_{LR}(i) } \simeq \abs{i\int_0^t d\tau \Tr \left\{\rho_{A}(0)\otimes\rho_{B}(0)\delta H_I(t -\tau-\beta_A \lambda/2,t -\tau-\beta_B \lambda/2)\right\}+\text{h.c.}}\notag\\
    \leq 2\int_0^t d\tau \norm{\delta H_I(t -\tau-\beta_A \lambda/2,t -\tau-\beta_B \lambda/2)}\times\norm{\rho_{A}(0)\otimes\rho_{B}(0)}.
\end{align}
In the second line, we have used the Cauchy-Schwartz inequality.  According to \cref{LRB}, we have $\abs{\delta G_{LR}(i) }=o(1)$ if $\abs{t+i\beta_{X}/2}<  \ell_{X}/v_{LR}$ holds for both $A$ and $B$.

According to \cref{NTAIE}, the error induced by truncation in \cref{TBHELAVG} can be calculated as
\begin{align}
    ( I| \mathcal{J}_{\rho_{A}(0)\otimes\rho_{B}(0)}^{1/2}  \circ [\mathcal{N}_{I}(i\beta/2)\circ\mathcal{U}^\dagger_{AB}(t)\circ\mathcal{N}^{\dagger}_{I}(-i\beta /2)-\mathcal{N}^{T}_{I}(i\beta /2)\circ\mathcal{U}^\dagger_{SE}(t)\circ\mathcal{N}^{T\dagger}_{I}(-i\beta /2)]|1)^\dagger\notag \\
     \simeq \int_0^{-\beta/2} d\tau_1( I| \mathcal{J}_{\rho_{A}(0)\otimes\rho_{B}(0)}^{1/2} \circ \left[\mathcal{L}^{\text{A}}_I(i\tau_1 )- \mathcal{L}^{\text{A},T}_I(i\tau_1 )\right]  \circ\mathcal{U}^\dagger_{b}(t)\circ[I+\int_0^{\beta/2} d\tau_2 \mathcal{L}^{\text{A}\dagger}_I(i\tau_2 )]|1)^\dagger\notag\\
     +\int_0^{\beta/2} d\tau_2( I| \mathcal{J}_{\rho_{A}(0)\otimes\rho_{B}(0)}^{1/2} \circ  [I+\int_0^{-\beta/2} d\tau_1\mathcal{L}^{\text{A},T}_I(i\tau_1 )] \circ\mathcal{U}^\dagger_{AB}(t)\circ\left[ \mathcal{L}^{\text{A}\dagger}_I(i\tau_2 )- \mathcal{L}^{\text{A},T\dagger}_I(i\tau_2 )\right]|1)^\dagger
 \end{align}
 Using this expression, we can constrain the difference 
 \begin{align}
     \abs{\delta G'_{LR}(i)}\simeq \abs{\int_0^{-\beta/2} d\tau_1 \Tr\left\{\rho_{A}(0)\otimes\rho_{B}(0)\delta H_I(i\tau_1)[\mathcal{U}^\dagger_{AB}(t)\circ(I+\int_0^{\beta/2} d\tau_2 \mathcal{L}^{\text{A}\dagger}_I(i\tau_2 ))I_{AB}]\right\}+\text{h.c.}}\notag\\
     +\abs{\int_0^{\beta/2} d\tau_2 \Tr\left\{[\mathcal{U}_{AB}(t)\circ(I+\int_0^{-\beta/2} d\tau_1\mathcal{L}^{\text{A},T}_I(i\tau_1 ))\rho_{A}(0)\otimes\rho_{B}(0)]\delta H_I(i\tau_2)I_{AB}\right\}+\text{h.c.}}\notag\\
     \leq 2\int_0^{\beta/2} d\tau \norm{\rho_{A}(0)\otimes\rho_{B}(0)}\times \norm{\delta H_I(i\tau)}\times \left[2\norm{I}+2\int_0^{\beta/2} d\tau'(\norm{H_I(i\tau')}+\norm{H^T_I(i\tau')})\right].
 \end{align}
 According to \cref{HIGRB,LRB}, we have $\abs{\delta G'_{LR}(i) }=o(1)$ if $\abs{i\beta/2}<  \ell_{X}/v_{LR}$ holds for both $A$ and $B$. Since $\abs{\overline{\delta G'_{LR}(i)} }\leq \overline{\abs{\delta G'_{LR}(i)} }=o(1)$,  the error is very small under long-time average.

\subsection{The error induced by deviation of the environment state in short-time regime}\label{EIDES}
The error induced by deviation of the environment state in \cref{EFNFT} equals to
\begin{equation}\label{GFT}
   ( I_{SE}|[\mathcal{J}_{\rho_{E}(0)}^{1/2}-\mathcal{J}_{\rho^\text{can}_E}^{1/2}] \circ \mathcal{J}_{\rho_S^\text{can}}^{1/2}\circ \mathcal{N}_\beta^{T\dagger}(t)\circ \mathcal{J}^{-1/2}_{\rho^\text{can}_S}|\rho_S(t))^\dagger=(O_{B_0}\otimes I_{\overline{B_0}}|\rho_E(0)-\rho^\text{can}_E)=(O_{B_0}|\Tr_{\overline{B_0}}[\rho_E(0)-\rho^\text{can}_E]),
\end{equation}
where $(O_{B_0}|=(\rho_S(t)\otimes \rho^\text{can}_{B_0}|\mathcal{U}^{T}_{SE}(t)\circ \mathcal{J}^{-1 /2}_{\rho^\text{can}_{B_0}}|I_S)$
is a $B_0$ local operator. For $\lambda=i$, using a similar approximation to \cref{UFOPA} for $\mathcal{U}^{T}_{SE}(t)$, we have
\begin{equation}\label{GRTOB}
    \norm{O_{B_0}}\simeq \norm{I_{B_0}+ \int_0^t d\tau (\rho_S(t)| \mathcal{U}^0_{S}(\tau) \circ\mathcal{J}^{1 /2}_{\rho^\text{can}_{B_0}}\circ \mathcal{L}_I\circ \mathcal{J}^{-1 /2}_{\rho^\text{can}_{B_0}}|I_S)} \leq \norm{I_{B_0}}+2 \int_0^t d\tau \norm{H_I(i\beta/2)}\times\norm{\rho_S(t)}.
\end{equation}
According to \cref{HIGRB}, we proceed to bound the above expression, yielding $\norm{O_{B_0}}=\Theta(1)$. It's worth mentioning that we employ the first-order perturbation approximation here, and higher-order terms can be bounded similarly. Furthermore, irrespective of the scenario, the operator $O_{B_0}$'s norm should not escalate with the size of the entire environment. Consequently, in line with \cref{DGETH}, we have:
\begin{equation}\label{ETHST}
    (O_{B_0}|\Tr_{\overline{B_0}}[\rho_E(0)-\rho^\text{can}_E])=O(N^{-\alpha}).
\end{equation}

The error induced by deviation of the environment state in \cref{NPFT} is
\begin{equation}
    (I_{AB}|[\mathcal{J}_{\rho_{A}(0)\otimes\rho_{B}(0)}^{1/2}-\mathcal{J}_{\rho^\text{can}_A\otimes\rho^\text{can}_B}^{1/2} ] \circ\mathcal{N}^{T\dagger}_{\beta_A,\beta_B}(t)|1)^\dagger =(O_{A_0B_0}|\Tr_{\overline{A_0}\overline{B_0}}[\rho_{A}(0)\otimes\rho_{B}(0)-\rho^\text{can}_A\otimes\rho^\text{can}_B]),
\end{equation}
where $(O_{A_0B_0}|=(\rho^\text{can}_{A_0}\otimes \rho^\text{can}_{B_0}|\mathcal{U}^{T}_{AB}(t)\circ \mathcal{J}^{-1 /2}_{\rho^\text{can}_{A_0}\otimes \rho^\text{can}_{B_0}}$ is an $A_0B_0$ local operator. Following the similar arguments as in \cref{ETHST}, in accordance with  \cref{DGETH}, we should have
\begin{equation}
    (O_{A_0B_0}|\Tr_{\overline{A_0}\overline{B_0}}[\rho_{A}(0)\otimes\rho_{B}(0)-\rho^\text{can}_A\otimes\rho^\text{can}_B]) =O(N_A^{-\alpha})+O(N_B^{-\alpha}),
\end{equation}
where $N_{X}$ is the size of the bath $X$.

\subsection{The error induced by deviation of the environment state in long-time regime}\label{EIDESLTR}
According to \cref{UELTA,TMLTR}, the term in \cref{FPETHLAVG} can be rewritten as
\begin{align}\label{FPERW}
   ( I_{SE}|[\mathcal{J}_{\rho_{E}(0)}^{1/2}-\mathcal{J}_{\rho^\text{can}_E}^{1/2}] \circ \mathcal{J}_{\rho_S^\text{can}}^{1/2}\circ \overline{\mathcal{N}_\beta^{T'\dagger}(t)}\circ \mathcal{J}^{-1/2}_{\rho^\text{can}_S}|\overline{\rho_S(t)})^\dagger\notag\\
   =\sum_a(\rho_S^\text{can}\otimes[\rho_{E}(0)-\rho^\text{can}_E]| \mathcal{N}^T_{I}(i\beta/2)|\Pi_a)^\dagger \times(\Pi_a|\mathcal{N}^{T\dagger}_{I}(-i\beta /2)\circ\mathcal{J}_{\rho_S^\text{can}}^{-1/2}|\overline{\rho_S(t)})^\dagger.
\end{align}
The energy width of $\rho_S^\text{can}\otimes\rho_{E}(0)$ and $\rho_S^\text{can}\otimes\rho^\text{can}_E$ are both narrower than $\Theta(N_{S+B}^{1/2})$ \cite{T18}. According to the theorem 2.1 in \cite{AKL16}, the superoperator $\mathcal{N}^T_{I}(-i\beta/2)$ at most increase the energy width by $\Theta(N_{B_0})$. According to the theorems 2.2 and 2.3 in \cite{AKL16}, the difference between the eigenstates of $H_0$ and $H$ causes the energy width to increase by $\Theta(\abs{H_I})$ at most. For the above reasons, the $a,b$ in \cref{FPERW,GTF2} can be limited to the energy shell $[E_c-\Delta E/2,E_c+\Delta E/2]$, where $\Delta E=\Theta(N_{B_0})+\Theta(N_{S+B}^{1/2})+\Theta(\abs{H_I})$ and $E_c=(\rho_S^\text{can}\otimes\rho_{E}(0)|H_0)$. The weights of eigenstates with energies above this range are suppressed exponentially. The energy width $\Delta E$ is much smaller than the average energy $\Theta(N_{SB})$. Their inverse temperature range given by \cref{ITFE} should be narrow and negligible in the thermodynamic limit. This means that it is difficult to distinguish these states from the same canonical ensemble $\rho_{SE}^\text{can}$ through local operator $ \mathcal{N}^{T\dagger}_{I}(-i\beta /2)\circ\mathcal{J}_{\rho_S^\text{can}}^{-1/2}|\overline{\rho_S(t)}\otimes I_E)$.  Therefore, we can do the following approximate replacement 
\begin{equation}\label{REP1}
    \sum_a |\Pi_a)(\Pi_a|\sim \sum_a |\Pi_a)(\rho_{SE}^\text{can}|=|I_{SE})(\rho_{SE}^\text{can}|,
\end{equation}
The error induced by replacement \cref{REP1} can be bounded as
\begin{align}\label{EIBRB}
    \abs{\sum_{a}(\rho_S^\text{can}\otimes[\rho_{E}(0)-\rho^\text{can}_E]| \mathcal{N}^T_{I}(i\beta/2)|\Pi_a)^\dagger \times(\Pi_a-\rho_{SE}^\text{can}|\mathcal{N}^{T\dagger}_{I}(-i\beta /2)\circ\mathcal{J}_{\rho_S^\text{can}}^{-1/2}|\overline{\rho_S(t)})^\dagger}\notag\\
    \leq \left[\abs{(\rho_S^\text{can}\otimes\rho_{E}(0)| \mathcal{N}^T_{I}(i\beta/2)|I)^\dagger} + \abs{(\rho_S^\text{can}\otimes\rho^\text{can}_E| \mathcal{N}^T_{I}(i\beta/2)|I)^\dagger}\right]\times \abs{(\Pi_{\max}-\rho_{SE}^\text{can}|\mathcal{N}^{T\dagger}_{I}(-i\beta /2)\circ\mathcal{J}_{\rho_S^\text{can}}^{-1/2}|\overline{\rho_S(t)})^\dagger},
 \end{align}
 where $E_{\max}\in [E_c-\Delta E/2,E_c+\Delta E/2]$ makes $\abs{(\Pi_{\max}-\rho_{SE}^\text{can}|\mathcal{N}^{T\dagger}_{I}(-i\beta /2)\circ\mathcal{J}_{\rho_S^\text{can}}^{-1/2}|\overline{\rho_S(t)})^\dagger}$  maximize. In \cref{EIBRB}, we use the properties that the  terms like $(\rho_S^\text{can}\otimes\rho_{E}| \mathcal{N}^T_{I}(i\beta/2)|\Pi_a)$ are positive.
  According to the previous analysis of the energy shell and \cref{DGETH}, \cref{EIBRB} can be bounded with $O(N^{-\alpha})$ and is small at the thermodynamic limit.  
The operator state $(I_S|\mathcal{J}_{\rho_S^\text{can}}^{1/2}\circ\mathcal{N}^T_{I}(i\beta/2)|I_{SE})$ also gives a $B_0$ local operator. After applying replacement (\ref{REP1}) to \cref{FPERW} and combining \cref{DGETH}, we have
\begin{equation}\label{BOGTF1}
    (I_S\otimes[\rho_{E}(0)-\rho^\text{can}_E]| \mathcal{J}_{\rho_S^\text{can}}^{1/2}\circ\mathcal{N}^T_{I}(i\beta/2)|I_{SE})^\dagger  \times(\rho_{SE}^\text{can}|\mathcal{N}^{T\dagger}_{I}(-i\beta /2)\circ\mathcal{J}_{\rho_S^\text{can}}^{-1/2}|\overline{\rho_S(t)})^\dagger=O(N^{-\alpha}).
\end{equation}

According to \cref{UELTA,TMLTR}, the error induced by the term in the second line of \cref{TBHELAVG} can be calculated with
 \begin{equation}\label{TDTLAVG}
    ( I_{AB}|[\mathcal{J}_{\rho_{A}(0)\otimes\rho_{B}(0)}^{1/2}-\mathcal{J}_{\rho^\text{can}_A\otimes\rho^\text{can}_B}^{1/2} ] \circ  \overline{\mathcal{N}^{T'\dagger}_{\beta}(t)}|1)^\dagger =\sum_a(\rho_{A}(0)\otimes\rho_{B}(0)-\rho^\text{can}_A\otimes\rho^\text{can}_B| \mathcal{N}^T_{I}(i\beta/2)|\Pi_a)^\dagger\times(\Pi_a|\mathcal{N}^{T\dagger}_{I}(-i\beta /2)|I_{AB})^\dagger
 \end{equation}
 Following the similar arguments in the preceding paragraph, the $a$ in \cref{TDTLAVG} can be limited to the energy shell $[E_{c'}-\Delta E/2,E_{c'}+\Delta E/2]$, where $\Delta E=\Theta(N_{B_0}+N_{A_0})+\Theta(N_{A+B}^{1/2})+\Theta(\abs{H_I})$ and $E_{c'}=(\rho_{A}(0)\otimes\rho_{B}(0)|H_0)$. The weights of eigenstates with energies above this range are suppressed exponentially. Moreover, since 
   $ \mathcal{N}^{T\dagger}_{I}(-i\beta /2)|I_{AB})=:|O_{A_0B_0}\otimes I_{\overline{A_0B_0}})$
gives an $A_0B_0$ local operator and the energy width to be considered here is also much smaller than the average energy $\Theta(N_{AB})$, the  ETH (\ref{DGETH}) allows us to replace $(\Pi_a|$ with  $(\rho_{AB}^\text{can}|$ and introduce only a small error. Now the approximate replacement becomes $\sum_a |\Pi_a)(\Pi_a|\sim \sum_a |\Pi_a)(\rho_{AB}^\text{can}|=|I_{AB})(\rho_{AB}^\text{can}|$.
The error induced by this replacement can be bounded in similar ways as \cref{EIBRB}. It is also small at the thermodynamic limit.
The operator state $(\rho_{A}(0)|\mathcal{N}^T_{I}(i\beta/2)|I_{AB})$ gives a $B_0$ local operator and $(\rho^\text{can}_{B}|\mathcal{N}^T_{I}(i\beta/2)|I_{AB})$ gives an $A_0$ local operator. So according to \cref{DGETH}, we have
\begin{equation}
    (\rho_{A}(0)\otimes\rho_{B}(0)-\rho^\text{can}_A\otimes\rho^\text{can}_B| \mathcal{N}^T_{I}(i\beta/2)|I_{AB})^\dagger \times(\rho_{AB}^\text{can}|\mathcal{N}^{T\dagger}_{I}(-i\beta /2)|I_{AB})^\dagger=O(N_{A}^{-\alpha})+O(N_{B}^{-\alpha}).
\end{equation}

\subsection{The error caused by time-dependent final state measurements}\label{ECEFS}
Now let us estimate the bound of \cref{GTF2}.
The energy width of $\rho_S(0)\otimes\rho_{E}(0)$  depend on the initial state, but it can be safely assumed to be narrower than $\Theta(N_S)$. After considering the difference between the eigenstates of $H_0$ and $H$, the  $a,b$ in \cref{GTF2} should also be in the energy shell $[E'_c-\Delta E'/2,E'_c+\Delta E'/2]$, where $\Delta E'=\Theta(N_S)+\Theta(\abs{H_I})$ and $E'_c=(H_0|\rho_S(0)\otimes\rho_E(0))$. The $a,b$ should be in the intersection of $[E_c-\Delta E/2,E_c+\Delta E/2]$ and $[E'-\Delta E'/2,E'+\Delta E'/2]$. The weights of eigenstates with energies above this range are suppressed exponentially. Since 
          $\mathcal{N}^{T\dagger}_{I}(-i\beta /2)\circ\mathcal{J}_{\rho_S^\text{can}}^{-1/2}|\Tr_E \Pi_{ab}\otimes I_E)$
also gives an $SB_0$ local operator. The off-diagonal ETH  (\ref{OFFDGETH}) tells
    \begin{equation}\label{ODETH}
       \abs{ (\Pi_{ab}| O_{SB_0}\otimes I_{\overline{B_0}})}= O(N^{-\alpha}) .
    \end{equation}
    The remaining part of \cref{GTF2} can be evaluated as follows
    \begin{align}\label{BOGF2}
        \sum_{\substack{a,b \\ a\neq b}} \big| (\rho_S^\text{can}\otimes \delta\rho_{E}|\mathcal{N}^T_{I}(i\beta/2)|\Pi_{ab})  (\Pi_{ab}|\rho_S(0)\otimes\rho_E(0)) \big|\notag\\
        \leq  \sum_{\substack{a,b}} \big| (\rho_S^\text{can}\otimes \delta\rho_{E}|\mathcal{N}^T_{I}(i\beta/2)|\Pi_{ab}) \sqrt{ (\Pi_{a}|\rho_S(0)\otimes\rho_E(0))}\sqrt{ (\Pi_{b}|\rho_S(0)\otimes\rho_E(0))} \big|\notag\\
        \leq  \norm{\abs{\mathcal{N}^{T\dag}_{I}(i\beta/2) (\rho_S^\text{can}\otimes\delta\rho_{E})}\sqrt{P}}_2\times \norm{\sqrt{P}}_2\leq \norm{\abs{\mathcal{N}^{T\dag}_{I}(i\beta/2) (\rho_S^\text{can}\otimes\delta\rho_{E})}} \leq 2 (1+2\int_0^{\beta/2} d\tau \norm{\abs{H_I(i\tau)}})=\Theta(1),
    \end{align}
    where the vector $\sqrt{P}=(\{\sqrt{ (\Pi_{a}|\rho_{SE}(0))}\})$, $\norm{\cdot}_2$ is vector 2-norm and $\abs{\cdot}$ replaces the matrix elements with their absolute values. In the second line, we have inserted the nonnegative part of $a=b$ and used $\abs{\rho_{ab}}^2\leq \rho_{a}\rho_{b} $. In the third line have used Cauchy-Schwartz inequality. Here we assume that the norm $\norm{\cdot}$ is consistent with  vector 2-norm. Additionally, we employ the approximation (\ref{NTAIE}) and \cref{HIGRB} in the third inequality sign on the third line.  Combining it with \cref{ODETH}, we can constrain  \cref{GTF2} to the following range
    \begin{equation}\label{BOGTF2}
         \Theta(1) \times O(N^{-\alpha}) .
    \end{equation}
\end{widetext}

\end{CJK*}

\end{document}